\documentclass[10pt]{article}
\usepackage{amsmath,amssymb,amsthm,cite,epsfig}
\usepackage{graphicx}
\usepackage{caption}
\usepackage{subcaption}
\usepackage{verbatim}
\usepackage{mathtools}
\usepackage[usenames, dvipsnames]{color}
\usepackage{hyperref} 

\usepackage{authblk}
\usepackage[toc,page]{appendix}
\numberwithin{equation}{section}

\def \beq{\begin{equation}}
\def \eeq{\end{equation}}
\def \beqa{\begin{eqnarray}}
\def \eeqa{\end{eqnarray}}

\def \l{\left(}
\def \r{\right)}

\begin{document}
\title{\begin{flushright}
\end{flushright}
{\bf Invariant ultraviolet scale corrections to the thermodynamics of degenerate Fermi gas and its implications}}
\author[1]{Dheeraj Kumar Mishra\thanks{dkmishra@imsc.res.in}}
\author[2]{Nitin Chandra\thanks{nitin.c.25@gmail.com}}
\affil[1]{The Institute of Mathematical Sciences, Chennai, Tamil Nadu, 600113, 
India.}
\affil[1]{Homi Bhabha National Institute,
Training School Complex, Anushakti Nagar, Mumbai, 400085, India}
\affil[2]{Department of Physics, National Institute of Technology,
Jamshedpur, 831014, India.}
\date{\empty}
\maketitle

\begin{abstract}
We study the invariant Planck scale correction to the thermodynamics of the ideal Fermi gas. We have considered the modified dispersion relation and the cut-off to the 
maximum possible momentum/energy (Planck energy) of the non-interacting ideal degenerate Fermi gas particles. 
With such a modification the expression for the degenerate pressure and the total energy gets modified accordingly.
We discuss the number density $n$ and mass $m$ dependence of the degenerate pressure.
We found that the degenerate pressure is perturbative in the SR limit which is quite unusual for a theory having an ultraviolet energy cut-off.
We then take the example of white dwarfs to explore the possible implications. Using this modified
degenerate pressure, we calculate the possible modification to the Chandrashekhar 
limit for white dwarfs using the Magueijo-Smolin (MS) modified dispersion
relation. The mass-radius M-R plot shows that the modified/corrected radius of the white dwarf can be  greater than, equal to and smaller than the usual special relativity (SR) value for particular masses.
We found that the Chandrasekhar mass limit gets a positive
correction i.e, the maximum possible mass for white dwarf increases in this formalism. 
We note that the presence of observed white dwarfs having radius smaller than the SR Chandrasekhar limit may find an explanation if they are modeled using a modified dispersion relation.
The correction, as stated before, is purely perturbative in the SR limit. Therefore this correction is solely because of the modified dispersion relation.
The value of the obtained degenerate pressure for a given mass is found to be greater than, equal to and smaller than the usual special relativity (SR) value for particular masses as expected.
It is shown by Mishra et al. that the Stefan-Boltzmann law gets a correction in such a theory with an ultraviolet cut-off.
Using this result we have calculated the luminosity of the white dwarf by taking
the model of partially
degenerate gas and considering the modified radiative envelope equation. In such
an analysis we observe that the pressure for a given mass and temperature value is less than that predicted by the usual SR theory. The luminosity also gets a negative correction. The correction to luminosity is nonperturbative as expected for such a theory.
\end{abstract}

\section{Introduction}
Modified Dispersion relation and effective theory with an ultraviolet cut-off are one of the aspects which gives us a possible way to explore beyond the known physics.
There has been many attempts to study the modification of the dispersion relation and its possible implications \cite{Rosati:2015pga, Husain:2015tna, Lobo:2016lxm, Gregg:2008jb, Sefiedgar:2010we, Ling:2009wj, Tao:2017mpe, Girelli:2006sc, Hinojosa:2015tga, Garattini:2011aa, Atteaga:2004}. 
Another aspect that has been explored in detail is the appearance and effect of an ultraviolet cut-off giving us an effective theory. The existence of such a cut-off is suggested by black hole physics \cite{Girelli:2005, Maggiore:1993rv, Park:2007az}. The cut-off is also predicted by almost all the candidate quantum gravity theories such as String theory, Loop Quantum Gravity and Non-commutative geometry and GUP (Generalized Uncertainity Principle) as well \cite{AmelinoCamelia:2000mn, Snyder:1946qz, Chandra:2014qva, Gross:1987ar, Amati:1988tn, Maggiore:1993kv, Garay:1994en, Kempf:1994su, Kempf:1996nk,
Smailagic:2003yb, Smailagic:2003rp, Kober:2010um, Nozari:2012gd}. We will especially consider the deformed relativistic theory which incorporates both the Modified Dispersion Relation (MDR) and an ultraviolet scale as cut-off. This scale must remain invariant in order to preserve the equivalence principle of relativity \cite{AmelinoCamelia:1997gz}\cite{AmelinoCamelia:2000mn}, otherwise different observers would see different scales for the same effective theory. 
This modifies the relativistic theory leading to so called Deformed/Doubly Special Relativity (DSR).
Along with the speed of light DSR incorporates this ultraviolet scale as an invariant energy/length scale in the relativistic theory. Magueijo and Smolin (MS) proposed a Lorentz algebra to incorporate the invariant scale such that the algebra remains intact but the representation becomes nonlinear. This in turn gives a
modified dispersion relation and puts an ultraviolet cut-off on the single particle
energy/momentum \cite{AmelinoCamelia:1997gz, AmelinoCamelia:2000mn, Magueijo:2001cr, Magueijo:2002am}.

In this article we will follow the MS
formalism \cite{Magueijo:2001cr}\cite{Magueijo:2002am} (also see for example
\cite{Chandra:2016wth} and \cite{Chandra:2011nj} ). Note that we have a modified relativistic theory with an invariant cut-off and MDR. This, in turn, means that the effects of such a modification can be observed not only at high energies but at low energies as well  \cite{Chandra:2016wth}\cite{Chandra:2011tf}\cite{Harms:2006dv}\cite{Santos:2018ler}. The thermodynamics of classical ideal gas and the gas of bosonic photons, in such a formalism, have also been studied in detail (see for example \cite{Chandra:2016wth}\cite{Chandra:2011nj}\cite{KowalskiGlikman:2001ct, Camacho:2007qy, Zhang:2011ms, Grether:2007ur}). In this respect studying Fermi gas becomes the next immediate thing. Since DSR gives us a modified dispersion relation and puts a cut-off on the highest single particle energy/momentum, this in effect will give a correction to the thermodynamics of the degenerate Fermi gas which becomes nontrivial.

As is well known, the model of degenerate Fermi gas is used to study the dynamics of many compact stars such as white dwarf stars, neutron stars etc. We will consider a simple model of white dwarf stars to study the possible implications of the obtained results. The white dwarf stars are the final stage of the stellar evolution after the nuclear processes inside the star have died down. Inside a white dwarf, which has used up almost all its fuel, practically no fusion is occurring. Therefore there is no source of thermal energy to support against the huge gravitational collapse. The stability, in this case, is
provided by what is known as degeneracy pressure which is the quantum pressure inside a degenerate Fermi gas. This idea was was suggested first by Fowler \cite{Fowler:1926} and Chandrasekhar \cite{Chandra:2005}. The degeneracy pressure at 0 K (considering relativistic quantum gas) is much
much larger than non-degenerate thermal pressure at very high density of stellar medium
(density keeps on increasing as the star keeps on collapsing under self
gravity) and it is this degeneracy pressure that supports the white dwarf against the
gravitational collapse.
We, therefore, can calculate the thermodynamic degenerate pressure
andequate it to the pressure due to gravity to get the mass and
radius relationship at equilibrium. This in turn gives the Chandrasekhar limit for white dwarfs in
the ultra-relativistic regime. Various stellar objects and their Chandrasekhar
mass limit have been studied in other such formalisms as well 
\cite{Camacho:2006qg}\cite{Gregg:2008jb}\cite{Bertolami:2009wa}\cite{
AmelinoCamelia:2009tv}. 
The effect on a compact star core is well studied in \cite{Wang:2011iv}.
\cite{Moussa:2014eda} and \cite{Moussa:2015yqy} study white dwarfs and their
Chandrasekhar limit in detail using GUP (Generalized Uncertainity Principle). 
\cite{Chandra:2016wth} studies in detail the modification of the equilibrium properties of blackbody radiation
in a theory with an ultraviolet cut-off using MS formalism. Amongst many results presented we note that
the Stefan-Boltzmann law gets modified in DSR and this result can be used to
study various stellar objects. 
It is a known fact that white dwarfs radiate 
much less than any other massive celestial white body and that too, is mainly a
surface phenomenon. 
The interior is completely degenerate but the surface, which radiates, is
non-degenerate matter.
Since the degenerate pressure and the density goes to zero at the
surface of the white dwarf, we have a thin envelope
of non-degenerate gas which is responsible for the radiation instead
of the whole bulk (see section 5.3 of \cite{paddy}).
This model of degenerate core with a radiative envelope of non-degenerate matter can then be used to calculate the luminosity of the star, which gets a negative and nonperturbative correction in DSR. We anticipate the nonperturbativity in SR limit due to the presence of ultraviolet cut-off (for details see \cite{Chandra:2016wth} \cite{Chandra:2011nj}).

Since we are considering massive Fermi particles, the study of the thermodynamics of such a massive particle gets separated in three different cases (for details see \cite{Chandra:2011nj}). 
In this paper we will first study the thermodynamics of degenerate Fermi gas with such a modified dispersion relation and an ultraviolet cut-off. 
We start by calculating the thermodynamic pressure and the total energy in all the three possible mass cases. We will especially look into the mass and number density dependence of the 
degenerate pressure in one of the cases in detail, both in SR ( as almost no literature discusses this dependence ) and the modified case. We will also briefly discuss the two extreme
nonrelativistic and the ultrarelativistic limits of the degenerate pressure.
The white dwarf is taken as an example to show one of the possible implications. The correction to the Chandrasekhar limit for white dwarf in all the three cases will be looked into.
Next, we calculate the luminosity-mass relationship for white dwarfs in both
the relativistic and the non-relativistic regimes separately. Finally, we will summarize the whole article and suggest works that may be done in future.

\section{Thermodynamics of degenerate Fermi gas}
The dispersion relation of a particle gets deformed in a relativistic theory
with an invariant ultraviolet energy scale.
In the MS model of the DSR \cite{Magueijo:2001cr}\cite{Magueijo:2002am} 
usual dispersion relation $E^2-p^2=m^2$ gets modified to
\begin{equation}
E^2-p^2=m^2\left(1-\frac{E}{\kappa}\right)^2
\label{MS}
\end{equation}
Here $E$, $p$ and $m$ are the total energy, magnitude of the 3-momentum and the
rest mass energy of the particle and
$\kappa$ is the invariant energy scale of the DSR theory ($\hbar=1,c=1$ and
$k_B=1$ unless or otherwise stated explicitly).
Note that the parameter $m$ is ``invariant mass" and is not the physical rest
mass of the particle.
To obtain the physical rest mass $m_0$, we put $p=0$ in the dispersion relation
(\ref{MS}). The dispersion relation (\ref{MS}) gives,
\begin{equation}\label{m}
m_0=\frac{m}{1+\frac{m}{\kappa}}\,\,\,\,\,\,\, \implies  m = \frac{m_0}{1-\frac{m_0}{\kappa}}
\end{equation}
where, $0\leq m_0\leq \kappa$ and $0\leq m\leq \infty$. For a detailed calculation see \cite{Chandra:2011nj}. Note that $m$ increases monotonically with increasing $m_0$.
We will now proceed to see the possible corrections to the Chandrasekhar limit of a white dwarf
in DSR.

We will start by considering a grand canonical ensemble of a degenerate Fermi gas
composed of $N$ relativistic electrons obeying Fermi-Dirac statistics. The total
number of particles can be calculated as (here the spin degeneracy $g=2s+1=2$),
\begin{equation} \label{TotalN}
 N =\int \int \langle n_p \rangle \frac{g d^3xd^3p}{h^3} = \frac{V_{ac}}{\pi^2}
\int_0^{\kappa} \frac{p^2 dp}{z^{-1}e^{\beta E(p)}+1}
\end{equation}
Here $\langle n_p \rangle$ is the mean occupation number and fugacity $z=e^{\beta \mu}$, $\mu$ being the chemical potential of the gas. For the Fermi gas
at $T=0K$, the mean occupation number is $\langle n_p \rangle=1$ for $E < \mu_0$
and $\langle n_p \rangle=0$ for $E > \mu_0$. Here $\mu_0=E_F$ is the
chemical potential at $T=0K$.
In such a case the number of particles become,
\begin{equation} \label{N}
 N = \int_{V_{ac}} \int_0^{p_F} \frac{d^3xd^3p}{h^3}.2 = \frac{V_{ac}
p_F^3}{3\pi^2}
\end{equation}
Here the accessible volume for a particle
$V_{ac}=V-\frac{4\pi}{3 \kappa^3}$ \cite{Chandra:2016wth}.
This gives the Fermi momentum,
\begin{equation} \label{p_F}
 p_F = (3\pi^2n)^{1/3}
\end{equation}
where $n=\frac{N}{V_{ac}}$ is the electron number density of the star.
The grand canonical partition function for such a gas is given by (see \cite{pathria}),
\begin{align}
 Q(V_{ac},T)=\prod_E \left(1+ze^{-\frac{E}{T}}\right)
\end{align}
The $q$-potential therefore becomes,
\begin{align}
 q\equiv\frac{P_{th}V_{ac}}{T}\equiv \ln Q(z,V_{ac},T)=\sum_E
\ln\left(1+ze^{-\frac{E}{T}}\right).
\end{align}
Taking the large volume limit we get,
\begin{align} 
 \bigg(\frac{P_{th}V_{ac}}{T}\bigg) = \frac{g V_{ac}}{(2\pi)^3} \int_{0}^{p_F}
\ln \left[1+ze^{-\beta E(p)}\right]{4\pi p^2 dp}
 \end{align}
Note that in such a large volume limit $V_{ac}=V-\frac{4\pi}{3 \kappa^3} \approx
V$.
The dispersion relation (\ref{MS}) for an electron gives
\begin{equation} \label{dE_dp}
 \frac{dE}{dp}=\frac{p}{\frac{m^2}{\kappa}+\left(1-\frac{m^2}{\kappa^2}\right)E}
\end{equation}
Using the above equation along with (\ref{TotalN}) we obtain,
 \begin{align}
    P_{th} &= \frac{T}{\pi^2}\left[\frac{p^3_F}{3}\ln \left[1+ze^{-\beta
E_F}\right]\right]+\frac{1}{3}n\left\langle p\frac{dE}{dp}\right\rangle
\nonumber \\
           &= \ln[1+ze^{-\beta E_F}] \left(\frac{NT}{V_{ac}}\right)
+\frac{1}{3}n\left\langle p\frac{dE}{dp}\right\rangle 
\end{align}
Since we are taking Fermi gas at $T=0K$, so the first term vanishes.
Using equations (\ref{N}) and (\ref{dE_dp}) and then changing the integration
variable to energy the expression of the thermodynamic pressure $P_{th}$ comes out to be,
\begin{align}\label{P_th}
 P_{th}  = \frac{1}{3}\, n\,\, \frac{\int_{V_{ac}} \int_0^{p_F}
\frac{d^3xd^3p}{(2 \pi)^3}.2.p\frac{dE}{dp}}{\int_{V_{ac}} \int_0^{p_F}
\frac{d^3xd^3p}{(2 \pi)^3}.2}= \frac{1}{3\pi^2} \int_{m_0}^{E_F}dE\,\,
\left[f(E)\right]^{3/2}
\end{align}
where $f(E)$ is given by,
\begin{equation} \label{fE}
f(E)=E^2-m^2\left(1-\frac{E}{\kappa}\right)^2 = (E-E_1)(E-E_2)
\end{equation}
with $E_1=m_0$ and
$E_2=\frac{m}{\frac{m}{\kappa}-1}=\frac{m_0}{\frac{2m_0}{\kappa}-1}$. The total energy is the next thing that we look into. The expression for total energy $U$ is given as,
\begin{align}
U= \sum_E E \langle n_E \rangle,
\end{align}
which in large volume limit for degenerate case becomes,
\begin{align}
U= \frac{V_{ac}}{\pi^2} \int_{0}^{p_F} p^2\,\, dp\,\, E
\end{align}
Expressing the above in terms of the energy $E$ we get,
\begin{align}\label{totalE}
U= \frac{V_{ac}}{\pi^2} \int_{m_0}^{E_F} E \left[E+
\frac{m^2}{\kappa}\l1-\frac{E}{\kappa}\r\right]
\left[E^2-m^2\l1-\frac{E}{\kappa}\r^2\right]^{\frac{1}{2}} dE
\end{align}

As stated before (i.e, in massive case) and also looking at the expressions of the integrals, we have three different cases namely (see for example \cite{Chandra:2011nj})
\begin{enumerate}
 \item $m_0<\frac{\kappa}{2} \Leftrightarrow m<\kappa$
\item $m_0=\frac{\kappa}{2} \Leftrightarrow m=\kappa$
\item $m_0>\frac{\kappa}{2} \Leftrightarrow m>\kappa$
\end{enumerate}
We will now consider each case separately.
\subsection{Case I: $m_0<\frac{\kappa}{2} \Leftrightarrow m<\kappa$}
In this case we have $-\infty<E_2<0<E_1=m_0<\frac{\kappa}{2}$.
The signature of $f(E)$, therefore, changes as follows
\begin{itemize}
 \item $f(E)$ is +ve for the regions $E<E_2$ and $E>E_1=m_0$
\item $f(E)$ is -ve for $E_2<E<E_1=m_0$
\end{itemize}
The integrand in equation (\ref{P_th}) remains real throughout the range of
integration and for this case we have
\begin{equation}
 [f(E)]^{3/2} = \left(1-\frac{m^2}{\kappa^2}\right)^{3/2}
\left[\left(E+\frac{m^2/\kappa}{1-\frac{m^2}{\kappa^2}}\right)^2-\frac{m^2}{
\left(1-\frac{m^2}{\kappa^2}\right)^2}\right]^{3/2}.
\end{equation}
The expression of pressure for this case becomes,
\begin{equation}\label{P_th_a}
 P_{th} = \frac{1}{3\pi^2} \left(1-\frac{m^2}{\kappa^2}\right)^{3/2}
\int_{m^\prime}^{E_F^\prime} dE^\prime (E^{\prime 2}-m^{\prime 2})^{3/2}.
\end{equation}
Note that $P_{th}$ is always positive, therefore we are considering the positive
root only. Here,
\begin{equation}\label{E_prime}
 E^\prime = E +\frac{m^2/\kappa}{1-\frac{m^2}{\kappa^2}}
\end{equation}
and
\begin{equation}\label{m_prime}
 m^\prime = \frac{m}{1-\frac{m^2}{\kappa^2}}
\end{equation}
We now change the variable to $x$ such that
 $E^\prime = m^\prime \cosh{x}$.
This change is perfectly allowed as we are considering the positive root only.
Note
that the limits of
$E^\prime$ ensures $E^\prime\geq m^\prime$ giving $\cosh{x} \geq 1$ (as $E_F$
can take values from $m_0$ to $\kappa$).
With the above substitution we have,
\begin{eqnarray}
 \int_{m^\prime}^{E_F^\prime} dE^\prime (E^{\prime 2}-m^{\prime 2})^{3/2} =
m^{\prime 4} \int_{0}^{x_F} dx\,\, \sinh^4x
= \frac{m^{\prime 4}}{4}
\left(\frac{1}{8}\sinh{4x_F}-\sinh{2x_F}+\frac{3}{2}x_F\right)
\label{int_E_prime}
\end{eqnarray}
Here we choose $x_F$ as the positive value of $x$ corresponding to the Fermi
energy ($E=E_F$), but choosing the negative value of $x$ will not change
the answer either.
Therefore the expression of the thermodynamic pressure becomes,
\begin{equation}\label{P_th_1}
 P_{th} = \frac{1}{3\pi^2} \left(1-\frac{m^2}{\kappa^2}\right)^{3/2} \frac{m^{\prime 4}}{4}
\left(\frac{1}{8}\sinh{4x_F}-\sinh{2x_F}+\frac{3}{2}x_F\right)
\end{equation}
Note that the dispersion relation (\ref{MS}) gives
\begin{equation}
 p_F^2 = f(E_F) = \left(1-\frac{m^2}{\kappa^2}\right)
\left[\left(E_F
+\frac{m^2/\kappa}{1-\frac{m^2}{\kappa^2}}\right)^2-\frac{m^2}{\left(1-\frac{m^2
}{\kappa^2}\right)^2}\right]
\end{equation}
The above equation leads to the value of Fermi energy as,
\begin{equation}\label{E_F}
E_F =
\frac{\sqrt{p_F^2\left(1-\frac{m^2}{\kappa^2}\right)+m^2}-\frac{m^2}{\kappa}}{
1-\frac{m^2}{\kappa^2}}
\end{equation}
which in turn gives,
\begin{equation}\label{cosh_x_F}
 \cosh{x_F} = \frac{E_F^\prime}{m^\prime} = \frac{\sqrt{p_F^2\left(1-\frac{m^2}{\kappa^2}\right)+m^2}}{m}
\end{equation}
For the familiar version, we will now express various expressions in terms of a new variable $z_F$ to give,
\begin{equation}
 \sinh{x_F} = \sqrt{\cosh^2x_F-1} = \frac{p_F}{m}\left(\sqrt{1-\frac{m^2}{\kappa^2}}\right)=z_F
\end{equation}
\begin{equation}
 \sinh{2x_F} = 2\sinh{x_F}\cosh{x_F} =
2z_F\sqrt{1+z_F^2}
\end{equation}
\begin{equation}\label{cosh_2x_F}
 \cosh{2x_F} = \sinh^2{x_F}+\cosh^2{x_F} = 1+2z_F^2
\end{equation}
\begin{equation}
 \sinh{4x_F} = 2\sinh{2x_F}\cosh{2x_F} =
4z_F\sqrt{1+z_F^2}\left(1+2z_F^2\right)
\end{equation}
Note that $x_F$, $\cosh{x_F}$, $\sinh{x_F}$, $\cosh{2x_F}$, $\sinh{2x_F}$,
$\sinh{4x_F}$, etc are all positive valued.
Putting the above expressions, the thermodynamic pressure (\ref{P_th_1}) becomes,
\begin{align}\label{P_th_DSR}
P_{th} = \frac{m^{\prime 4}}{24\pi^2}\left(1-\frac{m^2}{\kappa^2}\right)^{3/2}
 \left[z_F\sqrt{1+z_F^2}\left(2z_F^2-3\right)+ 3\ln\left[z_F+\sqrt{1+z_F^2}\right]\right] =\frac{m^{4}}{24\pi^2}\left(1-\frac{m^2}{\kappa^2}\right)^{-5/2} C(z_F)
 \end{align}
where $C(u)=\left[u\sqrt{1+u^2}\left(2u^2-3\right)+ 3\ln\left[u+\sqrt{1+u^2}\right]\right]$. 

\begin{figure}
\centering
\begin{subfigure}{.4\textwidth}
  \centering
  \includegraphics[width=1\linewidth]{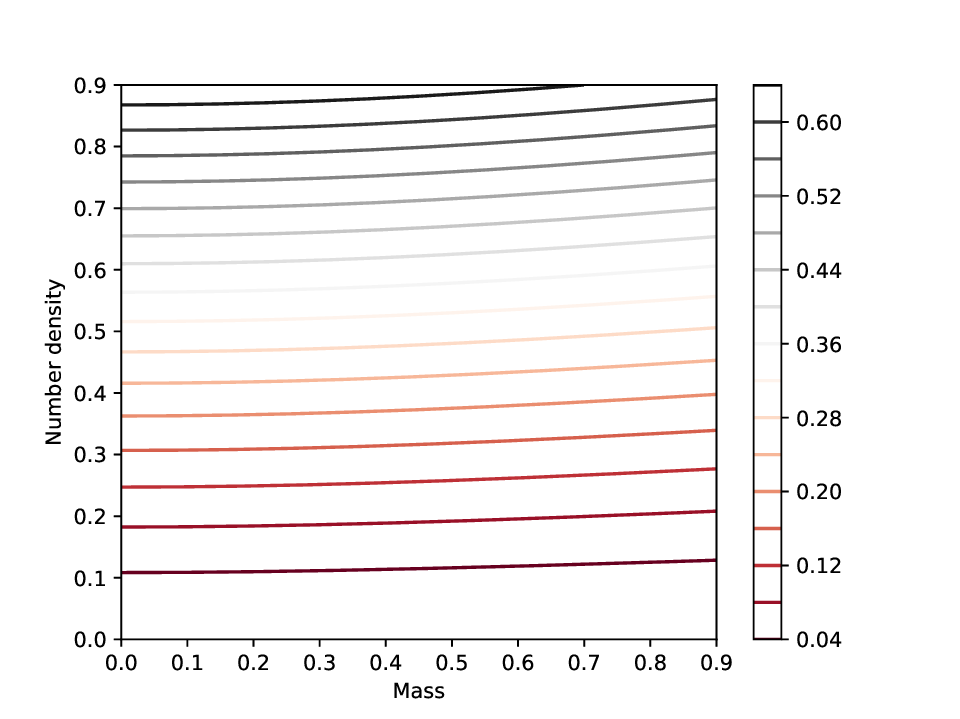}
  \caption{Usual SR Pressure Contour Plot}
  \label{fig:P}
\end{subfigure}
\begin{subfigure}{.4\textwidth}
  \centering
  \includegraphics[width=1\linewidth]{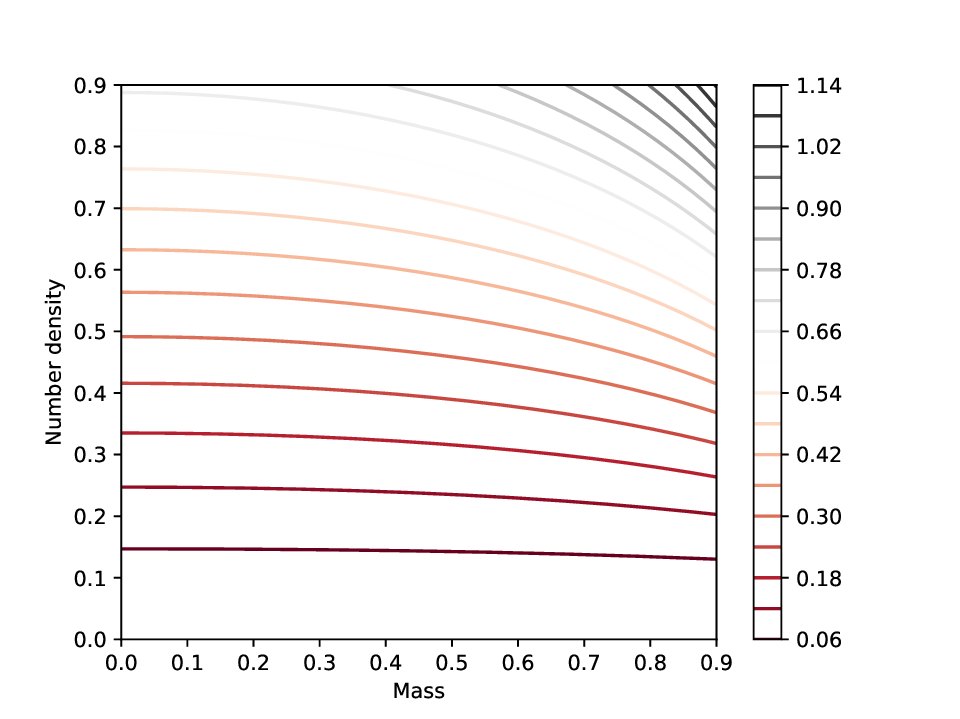}
  \caption{Modified Pressure Contour Plot}
  \label{fig:Pm}
\end{subfigure}
\begin{subfigure}{.4\textwidth}
  \centering
  \includegraphics[width=1\linewidth]{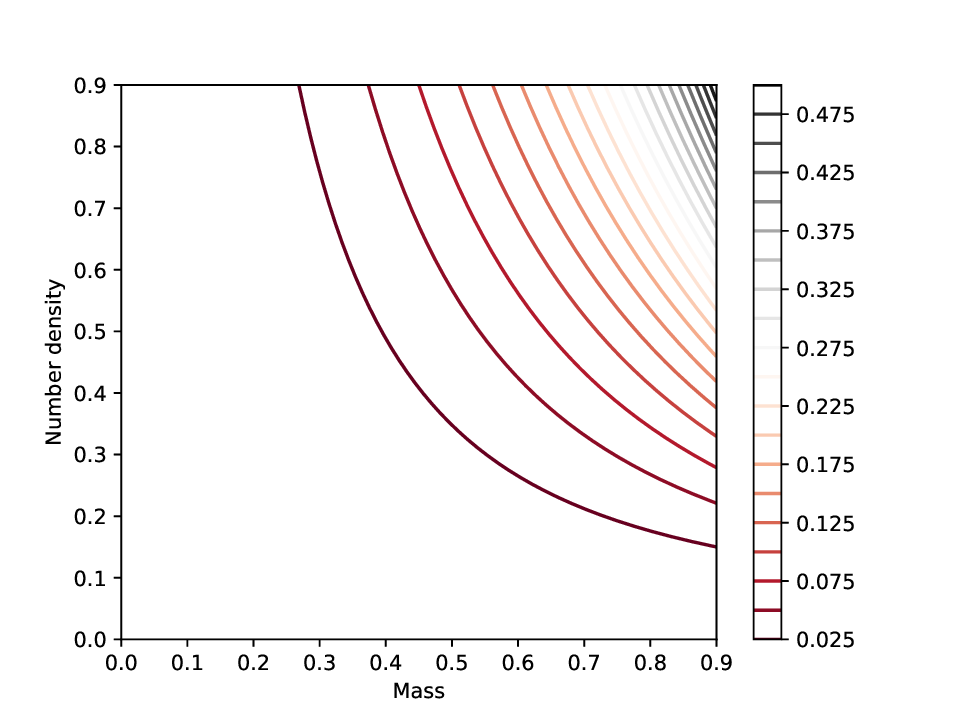}
  \caption{Difference of Modified and SR Pressure Contour Plot}
  \label{fig:PmP}
\end{subfigure}
\caption{\footnotesize{The plot (a) shows contour plot of degenerate pressure of the ideal Fermi gas with mass $m$ and number density $n$. The plot (b)
shows the modified pressure contour plot. The plot (c) shows the $(P_m- P)$ contour plot. Each colored line here represents a constant pressure value and 
we have chosen the values in Planck units with $\kappa=1$. We can clearly see that
the difference is positive for this particular range of the plots. Also the correction is more for larger $n$ and $m$ values.}}
\label{fig:modified}
\end{figure}

Note that the it is obvious from the expression of pressure (\ref{P_th_DSR}) that we have explicit dependence on three parameters namely modification parameter $\kappa$,
mass $m$ and number density $n$. The dependence of the degenerate pressure of Fermi gas on mass and number density is almost absent from the literature.
Therefore, for comparison let us, for this case explore the results in SR case as well. The degenerate pressure in SR case comes out to be,
\begin{align}
P^{SR}_{th} = \frac{m_0^{4}}{24\pi^2}
 \left[z\sqrt{1+z^2}\left(2z^2-3\right)+ 3\ln\left[z+\sqrt{1+z^2}\right]\right] =\frac{m_0^{4}}{24\pi^2} C(z)
\end{align}
here $z=\frac{(3\pi^2 n)^{1/3}}{m_0}$. 
We may now plot a contour map for the pressure in mass and number density. Note that each colored line in the contour plot
would represent the constant pressure value i.e, $P(m, n)= \text{Constant}$. 
Figure \ref{fig:P} and \ref{fig:Pm} shows the counter plots for the SR and the DSR cases for typical values of mass and number density in Planck units.
Note that the plot \ref{fig:PmP} shows that the correction is positive for this particular range of mass and number density.
But the correction can be positive or negative depending on the selected range of the $m$ and $n$. And another thing to note is that
the correction is more prominant for larger $n$ and $m$ values for the chosen scale $\kappa=1$.
Note that the value of the scale here is $\kappa=1$ for illustration, but can be appropriately be chosen for the given scenario.
Another important thing that we can now look into are the two extreme nonrelativistic and ultrarelativistic cases.
\subsubsection{$z<<1$: Non-relativistic SR case}
Using equation (4.6.31) of \cite{abramowitz}, the expression for the degenerate pressure becomes (keeping only the lowest orders),
\begin{align}
P^{SR}_{th} \approx \frac{m_0^{4}}{24\pi^2} \left(\frac{8}{5} z^5 -\frac{4}{7} z^7 \right) = \frac{(3 \pi)^{5/3}}{15 \pi^2} \frac{n^{5/3}}{m_0} -\frac{(3 \pi)^{7/3}}{42 \pi^2} \frac{n^{7/3}}{m_0^3}.  
\end{align}
This clearly implies the inverse dependence on the mass. 
\subsubsection{$z>>1$: Ultra-relativistic SR case}
Here again, using equation (4.6.31) of \cite{abramowitz}, the expression for the degenerate pressure becomes (keeping only the highest order),
\begin{align}
P^{SR}_{th} \approx \frac{m_0^{4}}{24\pi^2} 2 z^4 = \frac{(3 \pi)^{4/3}}{12\pi^2} n^{4/3}.  
\end{align}
Note that there is no mass dependence in this case.

We may now play the same game in DSR using the expressions (\ref{P_th_DSR}) and (\ref{cosh_x_F}). 
The corresponding expressions in DSR case are as follows,

\subsubsection{$z_F<<1$: Non-relativistic case}
The expression for the degenerate pressure becomes (keeping only the lowest orders),
\begin{align}
P_{th} \approx \frac{m^{4}}{24\pi^2} \left(1-\frac{m^2}{\kappa^2}\right)^{-5/2} \left(\frac{8}{5} z_F^5 -\frac{4}{7} z_F^7 \right) 
= \frac{(3 \pi)^{5/3}}{15 \pi^2} \frac{n^{5/3}}{m} -\frac{(3 \pi)^{7/3}}{42 \pi^2} \frac{n^{7/3}}{m^3} \left(1-\frac{m^2}{\kappa^2}\right) .  
\end{align}
Note that the correction comes in the next order.
\subsubsection{$z_F>>1$: Ultra-relativistic case}
The expression for the degenerate pressure becomes (keeping only the highest order),
\begin{align}
P_{th} \approx \frac{m^{4}}{24\pi^2} 2 \left(1-\frac{m^2}{\kappa^2}\right)^{-5/2} z_F^4 = \frac{(3 \pi)^{4/3}}{12 \pi^2} \left(1-\frac{m^2}{\kappa^2}\right)^{-1/2} n^{4/3}.  
\end{align}
Note that there is a mass dependence in this case as opposed to the SR one. It is obvious thar the SR limit gives the usual SR results given before.

Another very important point to note at this juncture is to see whether the 
expression for pressure (\ref{P_th_DSR}) is nonperturbative or perturbative in the SR limit as we expect the expression to be nonperturbative in 
theories with an ultraviolet cut-off (see for example \cite{Chandra:2011nj} and \cite{Chandra:2016wth} ). 
To see such a behavior we expand the above expression in $m^2/\kappa^2$ to get,
\begin{align}
P_{th}= P^{SR}_{th} + \frac{m^2}{\kappa^2}\left[\frac{5}{2} P^{SR}_{th} - \frac{m^{4}}{24\pi^2} \left\{ (1+u^2)^{-1/2} \left(u^4 +\frac{3}{2}u^3 +\frac{3}{2}u \right) 
+ (1+u^2)^{1/2} \left(3 u^2 +\frac{3}{2}u \right) \right\} \right] + \mathcal{O} \left(\frac{m^4}{\kappa^4}\right). 
\end{align}
This expression clearly shows that the the expression in the SR limit in perturbative as oppose to the expectation.
In typical DSR case the SR limit takes 
the upper limit of the integral over the single particle energy, describing the thermodynamic quantities, to infinity which is nonperturbative in nature.
On the other hand no such thing happens for a degenerate Fermi gas. 
The Fermi energy $E_F$ remains the upper limit of the integral in both SR and DSR. This explains the absence of nonperturbativity.

The expression of total energy $U$ can be similarly calculated using (\ref{totalE}) as,
\begin{align}
U &= \frac{V_{ac}}{\pi^2} \left(1-\frac{m^2}{\kappa^2}\right)^{5/2}  \int_{m^\prime}^{E_F^\prime} dE^\prime E^{\prime 2} (E^{\prime 2}-m^{\prime 2})^{1/2} - \frac{V_{ac}}{3\pi^2} \left(1-\frac{m^2}{\kappa^2}\right)^{3/2} \left(\frac{m^2}{\kappa}\right)  \int_{m^\prime}^{E_F^\prime} dE^\prime E^\prime (E^{\prime 2}-m^{\prime 2})^{1/2} \nonumber \\
&= \frac{V_{ac}}{\pi^2} \left(1-\frac{m^2}{\kappa^2}\right)^{5/2} m^{\prime 4} \left[ \int_{0}^{x_F} dx\,\, (\cosh \,x -1)\,\, \sinh^2x\,\, \cosh \,x \right] \nonumber\\
&=  \frac{V_{ac}}{\pi^2} \left(1-\frac{m^2}{\kappa^2}\right)^{5/2} m^{\prime 4} \left(\frac{1}{32}\sinh{4x_F}-\frac{1}{3}\sinh^3x_F-\frac{1}{8}x_F\right) \nonumber \\
&=  \frac{V_{ac}}{24\pi^2} \left(1-\frac{m^2}{\kappa^2}\right)^{5/2} m^{\prime 4} \left[8z_F^3\left(\sqrt{1+z_F^2}-1\right)-C(z_F)\right] =\frac{m^4 V_{ac}}{24\pi^2} \left(1-\frac{m^2}{\kappa^2}\right)^{-3/2}  D(z_F)
\end{align}
where  $D(u)=\left[8u^3\left(\sqrt{1+u^2}-1\right)-C(u)\right]$ and $C(u)$ is as defined above.

\subsection{Case II: $m_0=\frac{\kappa}{2} \Leftrightarrow m=\kappa$}
For this case the $f(E)$ becomes,
\begin{equation}
f(E)= 2\kappa \left(E-\frac{\kappa}{2}\right) 
\end{equation}
Putting the above in (\ref{P_th}) we get the expression of the pressure as
\begin{equation}
 P_{th} = \frac{2(2\kappa)^{3/2}}{15\pi^2}\left(E_F-\frac{\kappa}{2}\right)
^{5/2}
\end{equation}
Now using the dispersion relation (\ref{MS}) and (\ref{p_F}), in this case we have
\begin{equation}
 E_F-\frac{\kappa}{2} = \frac{p_F^2}{2\kappa}= \frac{(3\pi^2n)^{2/3}}{2\kappa}
\end{equation}
The expression of the total energy, therefore, in this case becomes,
\begin{align}\
U= \frac{\sqrt{2} V_{ac}\kappa^{3/2}}{3\pi^2}\left[ \frac{2}{5}\left(E_F-\frac{\kappa}{2}\right)^{5/2}- \frac{\kappa}{3}\left(E_F-\frac{\kappa}{2}\right)^{3/2}\right]
\end{align}

\subsection{Case III: $m_0>\frac{\kappa}{2} \Leftrightarrow m>\kappa$}
In this case we have $0<\frac{\kappa}{2}<E_1=m_0<\kappa<E_2<\infty$.
The signature of $f(E)$ therefore changes as follows
\begin{itemize}
 \item $f(E)$ is -ve for the regions $E<E_1=m_0$ and $E>E_2$
\item $f(E)$ is +ve for $m_0=E_1<E<E_2$
\end{itemize}
Again the integrand in equation (\ref{P_th}) remains real throughout the range
of integration and is given by
\begin{equation}
 [f(E)]^{3/2} = \left(\frac{m^2}{\kappa^2}-1\right)^{3/2}
\left[\frac{m^2}{\left(\frac{m^2}{\kappa^2}-1\right)^2}-\left(E-\frac{m^2/\kappa
}{\frac{m^2}{\kappa^2}-1}\right)^2\right]^{3/2}
\end{equation}
Thus the pressure therefore becomes
\begin{equation}
 P_{th} = \frac{1}{3\pi^2} \left(\frac{m^2}{\kappa^2}-1\right)^{3/2}
\displaystyle{\int_{-m^{\prime\prime}}^{-E_F^{\prime\prime}}} 
dE^\prime (m^{\prime\prime 2}-E^{\prime 2})^{3/2}
\end{equation}
where
\begin{equation}\label{E_F_2prime}
 E_F^{\prime\prime} = -E_F^\prime
\end{equation}
and
\begin{equation}\label{m_2prime}
 m^{\prime\prime} = -m^\prime,
\end{equation}
where $E_F^\prime$ is corresponding to (\ref{E_prime}) and $m^\prime$ is given
by
(\ref{m_prime}).
Here, $E_F^{\prime\prime}$ and $m^{\prime\prime}$ take only positive values and
$0\leq E_F^{\prime\prime} \leq m^{\prime\prime}$.
Now we again make the change of variable to $y$ such that
\begin{equation}
 E^\prime = -m^{\prime\prime} \cos{y}
\end{equation}
Limits of $E^\prime$ ensures that $0\geq E^\prime\geq -m^{\prime\prime}
\Rightarrow
0\leq \cos{y} \leq 1$.
Using this substitution we get,
\begin{eqnarray}
 \displaystyle{\int_{-m^{\prime\prime}}^{-E_F^{\prime\prime}}} dE^\prime
(m^{\prime\prime 2}-E^{\prime 2})^{3/2} 
= m^{\prime\prime 4} \int_{0}^{y_F} dy\,\, \sin^4y 
=\frac{m^{\prime\prime 4}}{4}
\left(\frac{1}{8}\sin{4y_F}-\sin{2y_F}+\frac{3}{2}y_F\right)
\label{int_E_2prime}
\end{eqnarray}
Here $y_F$ is the value of $y$ corresponding to the Fermi energy $E=E_F$ and
$0\leq y_F \leq \frac{\pi}{2}$.
The expression of the thermodynamic pressure therefore becomes,
\begin{equation}\label{P_th_3}
 P_{th} = \frac{1}{3\pi^2} \left(\frac{m^2}{\kappa^2}-1\right)^{3/2} \frac{m^{\prime\prime 4}}{4}
\left(\frac{1}{8}\sin{4y_F}-\sin{2y_F}+\frac{3}{2}y_F\right)
\end{equation}
Now, the dispersion relation (\ref{MS}) gives
\begin{equation}
 p_F^2 = f(E_F) = \left(\frac{m^2}{\kappa^2}-1\right)
\left[\frac{m^2}{\left(\frac{m^2}{\kappa^2}-1\right)^2}-\left(E_F-\frac{
m^2/\kappa}{\frac{m^2}{\kappa^2}-1}\right)^2\right].
\end{equation}
Rearranging the above equation we have,
\begin{equation}\label{E_F_prime_3}
E_F^\prime =
-\frac{\sqrt{m^2-p_F^2\left(\frac{m^2}{\kappa^2}-1\right)}}{\frac{m^2}{\kappa^2}
-1}
\end{equation}
Note that in this case $E_F^\prime$ is $-ve$.
The Fermi Energy $E_F$ is therefore given by
\begin{equation}\label{E_F_3}
E_F =
\frac{\frac{m^2}{\kappa}-\sqrt{m^2-p_F^2\left(\frac{m^2}{\kappa^2}-1\right)}}{
\frac{m^2}{\kappa^2}-1}.
\end{equation}
Note that for all values of $p_F \in [0,\kappa]$, the expression inside the
square-root is always positive. But the value of $p_F$ can at most be
$\kappa$ and in this case $m>\kappa$, therefore we only get the non-relativistic
particles or at most relativistic particles.
Physically this means that the particles are so heavy that their ultrarelativistic motion is not possible.
The above value of Fermi energy $E_F$ gives us,
\begin{equation}
 \cos{y_F} = -\frac{E_F^\prime}{m^{\prime\prime}} =\frac{\sqrt{m^2-p_F^2\left(\frac{m^2}{\kappa^2}-1\right)}}{m}
\end{equation}
As seen before, we will express the various quantities in terms of variable $q_F$ as,
\begin{equation}
 \sin{y_F} = \sqrt{1-\cos^2y_F} = q_F
\end{equation}
\begin{equation}
 \sin{2y_F} = 2\sin{y_F}\cos{y_F} =
2q_F\sqrt{1-q_F^2}
\end{equation}
\begin{equation}
 \cos{2y_F} = \cos^2{y_F}-\sin^2{y_F} = 1-2q_F^2
\end{equation}
\begin{equation}
 \sin{4y_F} = 2\sin{2y_F}\cos{2y_F} =
4q_F \sqrt{1-q_F^2}\left(1-2q_F^2\right)
\end{equation}
In this case also $y_F$, $\cos{y_F}$, $\sin{y_F}$, $\cos{2y_F}$, $\sin{2y_F}$,
$\sin{4y_F}$, etc are all positive valued.
Putting the above values in (\ref{int_E_2prime}) we get,
\begin{eqnarray}
&\displaystyle{\int_{-m^{\prime\prime}}^{-E_F^{\prime\prime}}} dE^\prime
(m^{\prime\prime 2}-E^{\prime 2})^{3/2}= \frac{m^{\prime\prime 4}}{8}
\left[3\sin^{-1}\left(q_F\right)-q_F\sqrt{1-q_F^2}\left(2q_F^2+3\right)\right]&
\end{eqnarray}
Hence, the thermodynamic pressure (\ref{P_th_3}) becomes,
\begin{equation}
P_{th} = \frac{m^{\prime\prime
4}}{24\pi^2}\left(\frac{m^2}{\kappa^2}-1\right)^{3/2}
\left[3\sin^{-1}\left(q_F\right)-q_F\sqrt{1-q_F^2}\left(2q_F^2+3\right)\right]= \frac{m^4}{24\pi^2}\left(\frac{m^2}{\kappa^2}-1\right)^{-5/2} J(q_F)
\end{equation}
where $J(u)=\left[3\sin^{-1}\left(u\right)-u\sqrt{1-u^2}\left(2u^2+3\right)\right]$.
Using (\ref{totalE}) the total energy in this case can be easily calculated as,
\begin{align}
U &= \frac{V_{ac}}{\pi^2} \left(\frac{m^2}{\kappa^2}-1\right)^{5/2}  \int_{-m^{\prime\prime}}^{-E_F^{\prime\prime}} dE^\prime E^{\prime 2} (m^{\prime\prime 2}-E^{\prime 2})^{1/2} + \frac{V_{ac}}{3\pi^2} \left(\frac{m^2}{\kappa^2}-1\right)^{3/2} \left(\frac{m^2}{\kappa}\right)   \int_{-m^{\prime\prime}}^{-E_F^{\prime\prime}} dE^\prime E^\prime (m^{\prime\prime 2}-E^{\prime 2})^{1/2} \nonumber \\
&= \frac{V_{ac}}{\pi^2} \left(\frac{m^2}{\kappa^2}-1\right)^{5/2} m^{\prime\prime 4} \left[ \int_{0}^{y_F} dx\,\, (\cos \,y -1)\,\, \sin^2y\,\, \cos \,y \right] \nonumber \\
&=  \frac{V_{ac}}{\pi^2} \left(\frac{m^2}{\kappa^2}-1\right)^{5/2} m^{\prime\prime 4} \left(-\frac{1}{32}\sin{4y_F}-\frac{1}{3}\sin^3y_F+\frac{1}{8}y_F\right) \nonumber \\
&=  \frac{V_{ac}}{24\pi^2} \left(\frac{m^2}{\kappa^2}-1\right)^{5/2} m^{\prime \prime4}\left[8q_F^3\left(\sqrt{1-q_F^2}-1\right)+C(q_F)\right] =\frac{m^4 V_{ac}}{24\pi^2} \left(1-\frac{m^2}{\kappa^2}\right)^{-3/2}  K(z_F)
\end{align}
where  $K(u)=\left[8u^3\left(\sqrt{1-u^2}-1\right)+J(u)\right]$ and $J(u)$ is as defined above.

The discussion till now is a modification in the thermodynamics of ideal relativistic Fermi gas in an effective theory with invariant ultraviolet cut-off. 
These results are valid at any energy depending on the choice of the of the parameters $n$, $m$ and the scale $\kappa$. One of the places which finds direct application of the degenerate 
Fermi gas is the study of the dynamics of the white dwarf stars. In the next section we explore the modified dynamics of such a typical white dwarf star as one of the examples of this formalism. 

\section{White Dwarfs: An Example}
A typical model of a white dwarf star consists of $N$ free electrons and
$\frac{N}{2}$ helium nuclei.
The mass of the star is given by ($m_n\simeq m_p$)
\begin{equation}\label{M}
 M = N m_e + \frac{N}{2} \left(2m_n+2m_p\right) \simeq  Nm_e +
2Nm_p=N(m_e+2m_p).
\end{equation}
Here $m_e$ and $m_p$ are the rest masses of the electron and the proton
respectively. 
In such stars the pressure support is given by the non-interacting (ideal) gas of
degenerate electrons and the mass density is mainly non-degenerate carbon or
helium ions \cite{paddy}\cite{paddy1}.
We may neglect the presence of the helium nuclei contribution to pressure as they do not contribute
significantly to the dynamics of the problem but only the mass.
The internal temperature of the white dwarf is of the order of $10^7$ K which is obviously not enough to hold the star against the self gravitational collapse.
It is easy to see that the Fermi energy of the electrons $E_F$ is of the order of $10^9$ K, which is higher than the average kinetic energy of the electron (~T). Hence the degeneracy condition holds and the gas of electrons can be approximated as a zero-temperature Fermi gas \cite{paddy}\cite{paddy1}\cite{stellar_astro}\cite{whitedwarf}. 
For a similar reason the effect of the radiation as well can be neglected as we choose $T=0 K$. Since all the
levels upto the Fermi level are filled (system is in the ground state)
and therefore there is no radiation effect showing up in the dynamics as a first approximation. 
Note, however, that it is an observed fact that the white dwarfs radiate and so have a finite luminosity that leads to its cooling. Therefore, they must have a thin radiative envelope along with a degenerate core.
Thus, the complete degenerate matter approximation is valid except at the thin non-degenerate surface envelope responsible for the finite luminosity of the white dwarf star. This we will discuss in detail in the luminosity section later in this article. 

We will take the star to be spherical in shape and therefore the change in the total
thermodynamic energy of the star ($dE_{th}$) 
due to infinitesimal change in the radius of the star ($dR$) is given by
\begin{equation}\label{E_th}
 dE_{th} = P_{th} dV = 4\pi R^2 P_{th} dR
\end{equation}
Whereas the change in the gravitational energy is given by,
\begin{equation}\label{E_g}
 dE_{g} = \alpha \frac{GM^2}{R^2}dR
\end{equation}
where $\alpha$ is of the order 1.
The exact value of $\alpha$ will depend on the spatial variation of $n$.
At equilibrium we have $dE_{th}= dE_g$, which leads to
\begin{equation}\label{PressureTh}
 P_{th} = \frac{\alpha G M^2}{4\pi R^4}
\end{equation}

\subsection{Case I: $m_0<\frac{\kappa}{2} \Leftrightarrow m<\kappa$}
Using (\ref{M}), (\ref{p_F}), (\ref{E_prime}) and (\ref{m_prime}) in (\ref{cosh_x_F})we get
\begin{equation}
 \cosh{x_F} = \sqrt{1+A^2\frac{M^{2/3}}{R^2}}
\end{equation}
where
\begin{equation}\label{A}
 A =
\frac{1}{\sqrt{mm^\prime}}\left(\frac{9\pi}{4m_e+8m_p}\right)^{1/3}=\frac{
\left(1-\frac{m_e}{\kappa} \right)\left[1-\left( \frac{m_e}{\kappa-m_e}\right)^2
\right]^{1/2}}{m_e}\left(\frac{9\pi}{4m_e+8m_p}\right)^{1/3}
\end{equation}
Note that here
$A\frac{M^{1/3}}{R}=\frac{p_F}{m}\left(\sqrt{1-\frac{m^2}{\kappa^2}}\right)$
gives relation between $A$ and $p_F$, which in the SR limit gives the correct
expected result. 
We will now express various expressions in terms of $A\frac{M^{1/3}}{R}$ to
give,
\begin{equation}
 \sinh{x_F} = \sqrt{\cosh^2x_F-1} = A\frac{M^{1/3}}{R}
\end{equation}
\begin{equation}
 \sinh{2x_F} = 2\sinh{x_F}\cosh{x_F} =
2A\frac{M^{1/3}}{R}\sqrt{1+A^2\frac{M^{2/3}}{R^2}}
\end{equation}
\begin{equation}
 \cosh{2x_F} = \sinh^2{x_F}+\cosh^2{x_F} = 1+2A^2\frac{M^{2/3}}{R^2}
\end{equation}
\begin{equation}
 \sinh{4x_F} = 2\sinh{2x_F}\cosh{2x_F} =
4A\frac{M^{1/3}}{R}\sqrt{1+A^2\frac{M^{2/3}}{R^2}}\left(1+2A^2\frac{M^{2/3}}{R^2
}\right)
\end{equation}
Putting the above expressions the thermodynamic pressure (\ref{P_th_1}) becomes,
\begin{equation}\label{P_th_1_full_expression}
P_{th} = \frac{m^{\prime 4}}{24\pi^2}\left(1-\frac{m^2}{\kappa^2}\right)^{3/2}
 \left[
A\frac{M^{1/3}}{R}\sqrt{1+A^2\frac{M^{2/3}}{R^2}}\left(2A^2\frac{M^{2/3}}{R^2}
-3\right)
+ 3\ln\left[\left(A\frac{M^{1/3}}{R}\right)+\sqrt{1+\left(A^2\frac{M^{2/3}}{R^2}
\right)}\right]\right]
\end{equation}
Equating equation (\ref{PressureTh}) with (\ref{P_th_1_full_expression}) we get the
mass-radius
relationship of the white dwarf star in DSR as,
\begin{align}
\frac{m^{\prime 4}}{24\pi^2}\left(1-\frac{m^2}{\kappa^2}\right)^{3/2} \left[
A\frac{M^{1/3}}{R}\sqrt{1+A^2\frac{M^{2/3}}{R^2}}\left(2A^2\frac{M^{2/3}}{R^2}
-3\right) +
3\ln\left[\left(A\frac{M^{1/3}}{R}\right)+\sqrt{1+\left(A^2\frac{M^{2/3}}{R^2}
\right)}\right]\right] 
= \frac{\alpha G M^2}{4\pi R^4} 
\end{align}
In the SR limit ($\kappa\rightarrow \infty$) we have $m^\prime\rightarrow m_e$
(see equation (\ref{m_prime})), 
$A\rightarrow \frac{1}{m_e}\left(\frac{9\pi}{4 m_e+ 8m_p}\right)^{1/3}$ (see
equation
(\ref{A})) and hence the above expression reduces to the 
correct relationship as given in section 8.5 of \cite{pathria}, with the
identification that $x=A \frac{M^{\frac{1}{3}}}{R}$.
\begin{figure}
\centering
\includegraphics[width=0.52\textwidth]{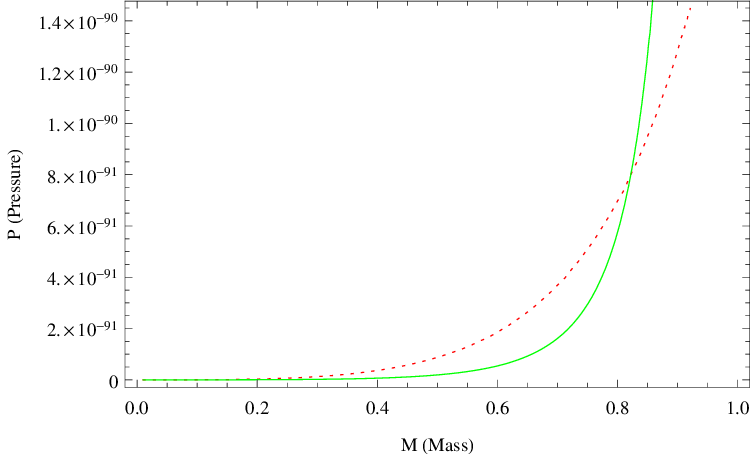}
\caption{Figure showing the variation of degeneracy pressure (P) with mass (M). Here also the mass is expressed in terms of the limiting mass $M_0^{SR}$ and the radius in terms of the Characteristic length of the order ${10}^{42}$ in Planck units. In this case too the value of the parameter $\kappa$ is $\kappa={10}^{-22}$. Here green line denotes the SR and the red dotted denotes the modified relation. Note that the value of the modified pressure is greater than the SR value for certain masses, equals and crosses to become less than the SR value for certain masses of the white dwarf as expected.}
\label{fig:Pplot}
\end{figure} 
\begin{figure}
\centering
\includegraphics[width=0.4\textwidth]{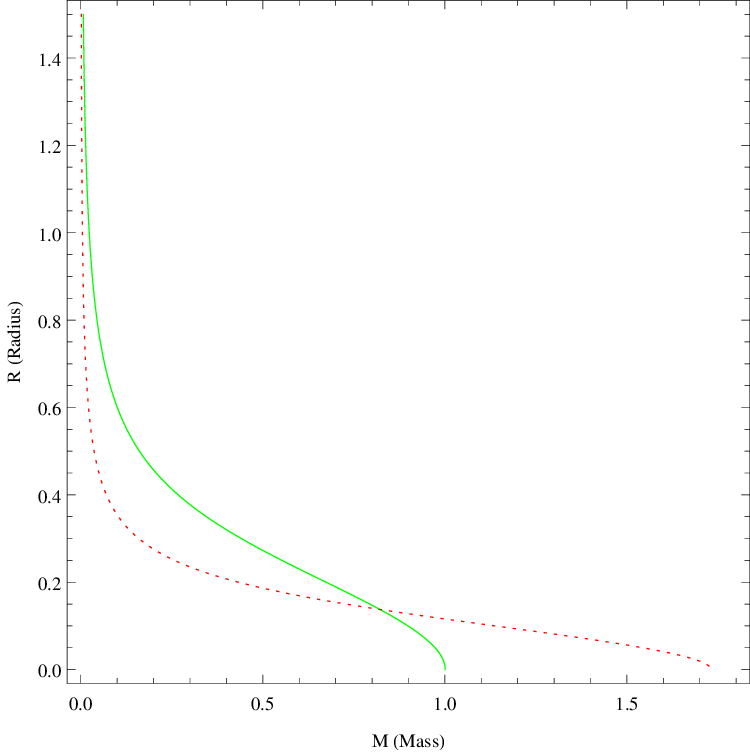}
\caption{Figure showing the Mass-Radius relationship. The mass is expressed in terms of the limiting mass $M_0^{SR}$ and the radius in terms of the Characteristic length of the order ${10}^{42}$ in Planck units. The value of the parameter $\kappa$ is $\kappa={10}^{-22}$. Here green line denotes the SR and the red dotted denotes the modified relation. Note that the modified M-R relation is below the SR one for some values of mass, equals the SR value at the crossing point and is greater than the SR value for certain masses of the white dwarfs. Note that the fact that the modified Chandrasekhar mass is greater than the SR mass is clearly visible from the plot.}
\label{fig:MRplot}
\end{figure} 
The M-R plot is shown in figure~\ref{fig:MRplot}. It is clearly obvious from figure~\ref{fig:MRplot} that the modified M-R relation is below the SR one for some values, equals the SR value at the crossing point and is greater than the SR value for certain masses of the white dwarfs. And figure~\ref{fig:Pplot} shows that the equilibrium degeneracy pressure is greater than the SR value for certain masses, equals and crosses to become less than the SR value for certain masses. This is easy to see as for a given mass value the thermodynamic pressure is inversely related to radius and since the radius for modified case is lesser than, equal to and greater than the SR case and so the modified pressure is expected to be greater than, equal to and less than corresponding the SR value. 
Since energy density is related directly to mass density as $\text{Mass} \,\, \text{Density}=\frac{\text{Energy}  \,\, \text{Density}}{c^2}$, therefore
denser compact objects are expected to show better measurable correction due to such a modification.
Note that in the plots, the scale $\kappa$ is chosen to be $\kappa={10}^{-22}$. 
Similar scale is noted by many who attempt to study the corrections to the dynamics of compact objects with modified dispersion relations \cite{Das:2008kaa} \cite{Ali:2011fa}\cite{Moussa:2015yqy}. This is the effective semi-classical scale where the effects starts to show up for considered white dwarf model. Note that this scale is of the order of the energy of the constituent particles as expected.
In future compact stars which are much denser can be studied where this scale might appropriately near the Planck scale. 
We will now look at both the non-relativistic and ultrarelativistic limits of the above obtained result.
\subsubsection{$R>>AM^{1/3}$: Non-relativistic case}
In such a limit, as stated before, using equation (4.6.31) of \cite{abramowitz}, 
the pressure (\ref{P_th_1_full_expression}) becomes,
\begin{align}
 P_{th} \simeq \frac{m^{\prime
4}}{15\pi^2}\left(1-\frac{m^2}{\kappa^2}\right)^{3/2}
\left(A^5\frac{M^{5/3}}{R^5}\right),
\end{align}
and the mass-radius
relationship becomes
\begin{equation}
 \frac{m^{\prime 4}A^5}{5}\frac{M^{5/3}}{R^5} \simeq \frac{3\pi \alpha G}{4}
\left(1-\frac{m^2}{\kappa^2}\right)^{-3/2} \frac{M^2}{R^4}
\end{equation}
This in turn gives
\begin{equation}
 R \simeq \frac{4m^{\prime 4}A^5}{15\pi\alpha
G}\left(1-\frac{m^2}{\kappa^2}\right)^{3/2}M^{-1/3}
\end{equation}
The SR limit gives the expected result (see for example equation (21) of section
8.5 of \cite{pathria}).
\subsubsection{$R<<AM^{1/3}$: Ultrarelativistic case
}
Here again using equation (4.6.31) of \cite{abramowitz}, in this limit 
the pressure (\ref{P_th_1_full_expression}) becomes
\begin{align}
 P_{th} \simeq \frac{ m^{\prime
4}}{12\pi^2}\left(1-\frac{m^2}{\kappa^2}\right)^{3/2} \left[
A^4\frac{M^{4/3}}{R^4}-A^2\frac{M^{2/3}}{R^2}\right],
\end{align}
and the relationship gives
\begin{equation}
 \frac{m^{\prime
4}A^4}{4}\frac{M^{4/3}}{R^4}\left(1-\frac{R^2}{A^2M^{2/3}}\right) \simeq 
\frac{3\pi \alpha G}{4} \left(1-\frac{m^2}{\kappa^2}\right)^{-3/2}
\frac{M^2}{R^4}
\end{equation}
We therefore have,
\begin{equation}
 R \simeq AM^{1/3}\left[1-\left(\frac{M}{M_0}\right)^{2/3}\right]^{1/2}
\end{equation}
where
\begin{equation}
 M_0 = \frac{(m^\prime A)^6}{(3\pi\alpha
G)^{3/2}}\left(1-\frac{m^2}{\kappa^2}\right)^{9/4}
\end{equation}
which is the modified Chandrashekhar mass limit, which is the maximum stable
mass for a white dwarf in this Chandrasekhar model. Using (\ref{m_prime}) and
(\ref{A})
the relation between SR and DSR case is,
\begin{equation}
M_0=\frac{M_0^{SR}}{\left[1-\frac{m^2}{\kappa^2}\right]^{\frac{3}{4}}}
\label{chandra_correct}
\end{equation}
Here also this gives the correct SR limit equations (see equation (22) and (23)
of
section 8.5 of \cite{pathria}) as expected. Since in this case $m<\kappa$, so
the denominator of the above equation is less than 1 , which implies that the
Chandrasekhar mass limit of the white dwarf has actually increased. This fact is clearly visible from the figure~\ref{fig:MRplot}. Interestingly, the SR limit of the above relation is purely perturbative and has no nonperturbative signature as was discussed before. 
The DSR correction to the Chandrasekhar limit comes solely because of the modification in the dispersion relation as the energy cut-off does not effect the calculation.
\subsection{Case II: $m_0=\frac{\kappa}{2} \Leftrightarrow m=\kappa$}
In this case, for a spherical model of star we have $n=\frac{N}{V}= \frac{3M}{8\pi m_p
R^3}$ and therefore the thermodynamic pressure becomes
\begin{equation}
 P_{th} =
\frac{1}{15\pi^{1/3}\kappa}\left(\frac{9}{4m_e+8m_p}\right)^{5/3}\frac{M^{5/3}}{
R^5}
\end{equation}
Equating the above to (\ref{PressureTh}) we get the mass-radius relationship as
\begin{equation}
 R = \frac{4}{15\pi\alpha\kappa
G}\left(\frac{9\pi}{4m_e+8m_p}\right)^{5/3}M^{-1/3}
\end{equation}
We conclude that there is no Chandrasekhar limit in this case for the masses of the white-dwarf stars.
\subsection{Case III: $m_0>\frac{\kappa}{2} \Leftrightarrow m>\kappa$}
Playing the same game, using (\ref{M}) and (\ref{p_F}) we have
\begin{equation}
 \cos{y_F} = -\frac{E_F^\prime}{m^{\prime\prime}} =
\sqrt{1-B^2\frac{M^{2/3}}{R^2}}
\end{equation}
where
\begin{equation}\label{B}
 B = \frac{1}{\sqrt{mm^{\prime\prime}}}\left(\frac{9\pi}{4m_e+8m_p}\right)^{1/3}
\end{equation}
As seen before, we will express the various quantities in terms of $B
\frac{M^{1/3}}{R}$ as,
\begin{equation}\label{sin_y_F}
 \sin{y_F} = \sqrt{1-\cos^2y_F} = B\frac{M^{1/3}}{R}
\end{equation}
\begin{equation}
 \sin{2y_F} = 2\sin{y_F}\cos{y_F} =
2B\frac{M^{1/3}}{R}\sqrt{1-B^2\frac{M^{2/3}}{R^2}}
\end{equation}
\begin{equation}
 \cos{2y_F} = \cos^2{y_F}-\sin^2{y_F} = 1-2B^2\frac{M^{2/3}}{R^2}
\end{equation}
\begin{equation}
 \sin{4y_F} = 2\sin{2y_F}\cos{2y_F} =
4B\frac{M^{1/3}}{R}\sqrt{1-B^2\frac{M^{2/3}}{R^2}}\left(1-2B^2\frac{M^{2/3}}{R^2
}\right)
\end{equation}
In this case also $y_F$, $\cos{y_F}$, $\sin{y_F}$, $\cos{2y_F}$, $\sin{2y_F}$,
$\sin{4y_F}$, etc are all positive valued.
Putting the above in (\ref{int_E_2prime})
\begin{eqnarray}
&\displaystyle{\int_{-m^{\prime\prime}}^{-E_F^{\prime\prime}}} dE^\prime
(m^{\prime\prime 2}-E^{\prime 2})^{3/2}= \frac{m^{\prime\prime 4}}{8}
\left[3\sin^{-1}\left(B\frac{M^{1/3}}{R}\right)-B\frac{M^{1/3}}{R}\sqrt{
1-B^2\frac{M^{2/3}}{R^2}}\left(2B^2\frac{M^{2/3}}{R^2}+3\right)\right]&
\end{eqnarray}
Hence, the thermodynamic pressure (\ref{P_th_3}) is given by
\begin{equation}\label{P_th_3_full_expression}
P_{th} = \frac{m^{\prime\prime
4}}{24\pi^2}\left(\frac{m^2}{\kappa^2}-1\right)^{3/2}
\left[3\sin^{-1}\left(B\frac{M^{1/3}}{R}\right)
-B\frac{M^{1/3}}{R}\sqrt{1-B^2\frac{M^{2/3}}{R^2}}\left(2B^2\frac{M^{2/3}}{R^2}
+3\right)\right]
\end{equation}

Now at equilibrium we equate the above expression of pressure to
(\ref{PressureTh}) and get the mass-radius relationship of the white dwarf star
as,
\begin{align}
 \frac{m^{\prime\prime 4}}{24\pi^2}\left(\frac{m^2}{\kappa^2}-1\right)^{3/2}
\left[3\sin^{-1}\left(B\frac{M^{1/3}}{R}\right)-B\frac{M^{1/3}}{R}\sqrt{
1-B^2\frac{M^{2/3}}{R^2}}\left(2B^2\frac{M^{2/3}}{R^2}+3\right)\right]
     = \frac{\alpha G M^2}{4\pi R^4} 
\end{align}
The equation (\ref{sin_y_F}) tells us that $0\leq B\frac{M^{1/3}}{R} \leq 1
\Rightarrow
R\geq BM^{1/3}$ as $0\leq y_F \leq \frac{\pi}{2}$.
The asymptotic behavior $R>>BM^{1/3}$ of the mass-radius relationship, using
equation (4.4.40) of \cite{abramowitz}, is given by
\begin{equation}
 R \simeq \frac{4m^{\prime\prime 4}B^5}{15\pi\alpha
G}\left(\frac{m^2}{\kappa^2}-1\right)^{3/2}M^{-1/3}
\end{equation}
Again, there is no limit for the masses of the white-dwarf stars as expected.
Note that the Chandrasekhar limit only in ultrarelativistic case which is
not possible.

We conclude this section by noting that we only get the Chandrasekhar limit in
ultrarelativistic limit of the $m<\kappa$ case, which apparently is the physical
case. We also note that this correction is actually positive.
Therefore the mass- radius relation has changed and so the radius of white dwarf
is lower, equal to and greater than the SR values for given masses. The decrease in radius is theoretically predicted \cite{Moussa:2014eda}\cite{
Moussa:2015yqy} and experimentally observed
as well \cite{max}\cite{Panei:1999ji}\cite{Provencal:2002}\cite{Mathews:2006nq}. This analysis may not be the only explanation of the observed decrease but is surely an attempt. In future we may observe the white dwarfs with radius greater than that predicted by the present SR theory.
\section{Modified Structure equations}
In previous sections we considered the matter density to be constant.
In this section we will consider variable matter density and obtain the static
structure of the white dwarf using
stellar structure equations \cite{whitedwarf}. As is obvious from the analysis of the previous section that
the Chandrasekhar limit is obtained only for case $I$ i.e., $m<\kappa$. We will, therefore, consider that case only for the present analysis.
\subsection{Lane-Emden equation and the Chandrasekhar mass}
For the stellar model we first need to express the expressions in terms of the
matter density $\rho$ instead of number density.
We will use  the mean molecular weight $\mu=\frac{\rho}{n m_H}$,
where $\frac{1}{m_H}=N_A$  as the gas constant (Avogadro's Number), $m_H=1.6605\times 10^{-27}$ Kg is atomic mass constant (equal to mass of hydrogen atom for all practical purposes) and $n$ is number
density as usual. Therefore, for $m<\kappa$ case the Fermi momentum is given as,
\begin{align}\label{fermi_p}
p_F&=\left( A\frac{M^{1/3}}{R} \right)\frac{m}{\sqrt{1-\frac{m^2}{\kappa^2}}}\nonumber \\
&=(3 \pi^2 n)^{1/3}=\left(3 \pi^2 \frac{\rho}{\mu m_H}\right)^{1/3}
\end{align}
The pressure in nonrelativistic case becomes,
\begin{align}
 P_{th} \simeq \frac{m^{\prime
4}}{15\pi^2}\left(1-\frac{m^2}{\kappa^2}\right)^{3/2}
\left(A^5\frac{M^{5/3}}{R^5}\right)=C_1 \rho^{5/3},
\end{align}
where $C_1=\frac{\left(3\pi^2\right)^{2/3}}{5m} \frac{1}{(\mu m_H)^{5/3}}$.
Similarly, the pressure in ultrarelativistic case becomes,
\begin{align}
 P_{th} \simeq \frac{ m^{\prime
4}}{12\pi^2}\left(1-\frac{m^2}{\kappa^2}\right)^{3/2} \left(
A^4\frac{M^{4/3}}{R^4}\right)=C_2 \rho^{4/3},
\end{align}
where $C_2=\frac{1}{4}\left(
1-\frac{m^2}{\kappa^2}\right)^{-1/2}(3\pi^2)^{1/3}\frac{1}{(\mu m_H)^{4/3}}$.
Therefore, we conclude that both in non-relativistic and ultrarelativistic case
the degenerate pressure depends on the density of matter and the degenerate
electron gas behaves as a perfect gas with the polytropic equation of state.
Then assuming the mass density only depends on pressure, not on temperature,
we can solve the structure equations. 
The general polytrope is of the form,
\begin{align}\label{Ppolytrope}
 P=K \rho^{\gamma}
\end{align}
where $\gamma=1+\frac{1}{n}$ and $n$ is called polytropic index. 
Note that here the $\kappa$ dependence is in the coefficient $K$.
We will then use this in the Poisson equation to
get the usual Lane-Emden equation and which further gives the expressions for the radius $R$ and mass
$M$.
For a non-rotating fluid we have the hydrostatic equilibrium structure equations
as (see \cite{paddy}\cite{paddy1}\cite{whitedwarf}),
\begin{align}
 \frac{d m(r)}{d r}=4\pi r^2 \rho
\end{align}
and 
\begin{align}
 \frac{d P}{d r}=-\frac{G m(r) \rho}{r^2}.
\end{align}
Combining the above two equations we get,
\begin{align}\label{DifferentialP}
 \frac{1}{r^2}\frac{d}{d r}\left( \frac{r^2}{\rho} \frac{d P}{d r}\right)=-4\pi
G\rho
\end{align}
Using (\ref{Ppolytrope}) and the boundary conditions $\rho(r=0)=\rho_0$ and
$\frac{d \rho}{d r}\big|_{(r=0)}=0$, we can obtain $\rho(r)$ by solving
(\ref{DifferentialP}).
We will, instead, scale this to the dimensionless form using
\begin{align}\label{Scale}
\rho(r)=\rho_0 \theta^n (r) \,\,\,\, \text{and} \,\,\,\, r=\left(\frac{(n+1) K
{\rho_0}^{(1-n)/n}}{4 \pi G} \right)^{1/2} \xi= a \xi, 
\end{align}
here $\theta$ and $\xi$ are the dimensionless
density and radius respectively and $a$ is the scale factor.
With the above substitution in (\ref{DifferentialP}) we get the Lane-Emden
equation for polytrope index $n$ as,
\begin{align}\label{DifferentialTheta}
 \frac{1}{\xi^2}\frac{d}{d \xi}\left( \xi^2 \frac{d \theta}{d
\xi}\right)=-\theta^n.
\end{align}
\begin{figure}
\centering
\includegraphics[width=0.8\textwidth]{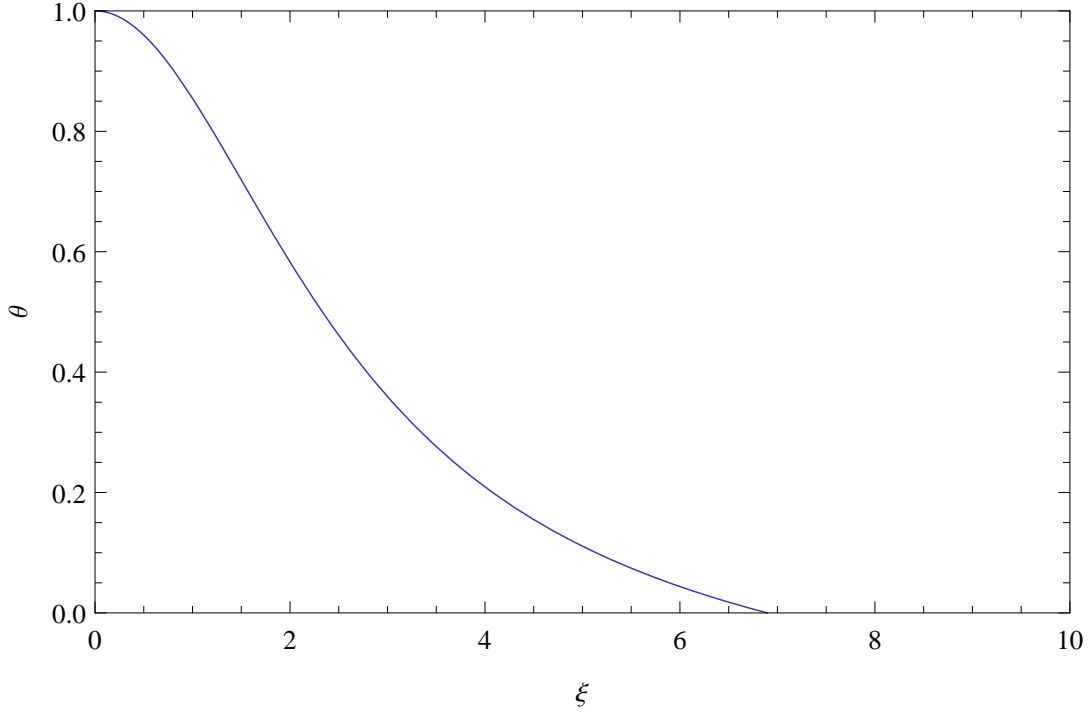}
\caption{Figure showing the numerical solution of (\ref{DifferentialTheta}) for
$n=3$, i.e. the variation of $\theta(\xi)$ as a function of $\xi$. It can be
clearly seen from the plot that the value of $\xi_1=6.89$
corresponding to $\theta=0$.}
\label{fig:DiffSol}
\end{figure} 
Using the boundary conditions $\theta(r=0)=1$ and $\frac{d \theta}{d
r}\big|_{r=0}=0$, the above equation can be integrated numerically, starting
from $\xi=0$ for a particular choice of $\kappa$. 
One can see that for $n<5$, the solutions decrease monotonically with the zero
at $\xi=\xi_1$ i.e. $\theta(\xi_1)=0$ or $\rho(r_1=a\xi_1)=0$. This
$r_1=a\xi_1=R$ is the radius of the star,
\begin{align}
 R=\left(\frac{(n+1) K {\rho_0}^{(1-n)/n}}{4
\pi G} \right)^{1/2} \xi_1
\end{align}
Using (\ref{Scale}) and (\ref{DifferentialTheta}) we get,
\begin{align}
 M=\int_0^R 4\pi r^2 \rho dr =4\pi \left(\frac{(n+1) K}{4
\pi G} \right)^{3/2} {\rho_0}^{(3-n)/2n} \xi_1^2|\theta^{\prime}(\xi_1)|.
\end{align}
We get the mass-radius relation  by eliminating the $\rho_0$ between $R$ and $M$
relation as,
\begin{align}
M=4\pi \left(\frac{(n+1) K}{4 \pi G} \right)^{n/(n+1)} \xi_1^{(n+3)/(1-n)}
\xi_1^2|\theta^{\prime}(\xi_1)| R^{(3-n)/(1-n)}. 
\end{align}
The interesting case is the ultrarelativistic case with $\gamma=\frac{4}{3}$ or
$n=3$ and
the corresponding values are $\xi_1=6.89$ and
$\xi^2|\theta^{\prime}(\xi_1)|=2.02$ (refer to \cite{Chandra:1931}). The numerical
solution is plotted in figure~\ref{fig:DiffSol}.
The mass then becomes,
\begin{align}
 M_0 = \left[ \frac{(3\pi)^{1/2}}{{m_H}^2 G^2} \left(
1-\frac{m^2}{\kappa^2}\right)^{-3/4} \right]
\left( \frac{2.02}{2
\mu^2}\right)=\frac{M_0^{SR}}{\left[1-\frac{m^2}{\kappa^2} \right]^{3/4}}
\end{align}
which is exactly the mass $M_0$ obtained in section 3.1.2 above, with the proper
substitutions. Note that the white dwarfs which are predominantly made up of
$^{12}C$ or $^{16}O$, the value of $\mu\simeq2$. The value can be written in
terms of
the mass of the sun with proper units introduced as,
\begin{align}
 M_0=\frac{5.7}{ \mu^2 \left[1-\frac{m^2}{\kappa^2} \right]^{3/4}} M_\odot,
\end{align}
where $M_\odot$ is the mass of the sun. Note that this mass is independent of
radius $R$ and the central density $\rho_0$.
The radius can then be written as,
\begin{align}
 R= \left[ \frac{(3\pi)^{1/2}}{m_e m_H \sqrt{G}} \left(
1-\frac{m^2}{\kappa^2}\right)^{-1/4} \right]
\left( \frac{6.89}{2 \mu}\right)
\left(\frac{\rho_c}{\rho_0}\right)^{1/3}
\end{align}
where $\rho_c=\frac{m_H \mu {m_e}^3}{3 \pi^2}$ is the critical density which defines a rough partition between non-relativistic and and relativistic regimes. Such that, $\rho<<\rho_c$ corresponds to non-relativistic case and $\rho>>\rho_c$ corresponds to ultrarelativistic case.
\subsection{General structure equation}
In the previous section we assumed a particular polytropic form, with polytropic
index $n$, of the density dependence of pressure. 
The non-relativistic and ultrarelativistic cases can be understood by taking a
particular values of $n$ and solving the differential equation.
But we will now take a more general approach than just taking the two 
extreme cases.  
We first express the number density in terms of matter density as $n=\frac{\rho}{\mu m_H}$ in (\ref{fermi_p}) to give the expression of density as,
\begin{align}
 \rho= \frac{\mu m_H m^3}{3 \pi^2} \left( 1-\frac{m^2}{\kappa^2}\right)^{-3/2}
A^3\frac{M}{R^3}=C_1 \mu x^3 
\end{align}
where $C_1=\frac{m_H m^3}{3 \pi^2} \left(
1-\frac{m^2}{\kappa^2}\right)^{-3/2}=\frac{m_H m_e^3}{3 \pi^2\left(
1-\frac{m_e}{\kappa}\right)^3} \left[
1-\left(\frac{m_e}{\kappa-m_e}\right)^2\right]^{-3/2}$ and 
$x=A\frac{M^{1/3}}{R}$ i.e, \\
$x=\left[\frac{
\left(1-\frac{m_e}{\kappa} \right)\left[1-\left( \frac{m_e}{\kappa-m_e}\right)^2
\right]^{1/2}}{m_e}\left(\frac{9\pi}{4m_e+8m_p}\right)^{1/3}\right]\frac{M^{1/3}
}{R}$.
The pressure is given by (\ref{P_th_1_full_expression})
as, 
\begin{align}
P_{th}=C_2 F(x),
\end{align}
where
$C_2=\frac{m^4}{24\pi^2}\left(1-\frac{m^2}{\kappa^2}\right)^{-5/2}=\frac{m_e^4}{
24 \pi^2\left(
1-\frac{m_e}{\kappa}\right)^4} \left[
1-\left(\frac{m_e}{\kappa-m_e}\right)^2\right]^{-5/2}$
and $F(x)=x(2x^2-3)(1+x^2)^{1/2}+ 3\sinh^{-1}(x)$.
Note that the dependence of $\kappa$ is in the coefficients $C_1$ and $C_2$.
The spherically symmetric fluid in equilibrium with gravitational force is given
by (\ref{DifferentialP}). In order to cast it into a convenient form, we will
make a change of variable as $z^2\equiv (x^2+1)$ and find that
\begin{align}
 \frac{1}{\rho} \frac{d P}{d r}= \frac{8 C_2}{C_1 \mu}\left(\frac{d z}{d
r}\right)
\end{align}
Using the above equation in (\ref{DifferentialP}) we get,
\begin{align}\label{DifferentialQ}
 \frac{1}{r^2}\frac{d}{dr}\left(r^2 \frac{d z}{d r} \right)=-\frac{\pi}{2}
\left(\frac{G{C_1}^2 \mu^2}{C_2}\right)(z^2-1)^{3/2}
\end{align}
Let us take the value of $z$ at $r=0$ as $z_c$. As was done previously, we will
rescale the variables as 
\begin{align}\label{Scale2}
z=Q z_c \,\,\,\, \text{and} \,\,\,\, r=\sqrt{\frac{2 C_2}{\pi G}}\left(
\frac{1}{C_1 \mu z_c}\right) \xi=a \xi. 
\end{align}
Using the above scaling (\ref{DifferentialQ}) can be written in terms of new
variables $Q$ and $\xi$ as,
\begin{align}
 \frac{dQ^2}{d\xi^2}+\frac{2}{\xi} \frac{d Q}{d \xi}+\left(Q^2-\frac{1}{z_c^2}
\right)^{3/2}=0
\end{align}
with the boundary conditions $Q(\xi=0)=1$ (by definition) and $\frac{d Q}{d
\xi}\big|_{\xi=0}=0$ (by assuming that the gradient of pressure at the origin
vanishes). Given $z_c>1$, this can be numerically solved for the given boundary
conditions outwards from $\xi=0$. The density can as well be written in terms of
the new variables as,
\begin{align}
 \rho=C_1 \mu x^3=C_1\mu(z^2-1)^{3/2}=C_1 \mu z_c^3 \left(
Q^2-\frac{1}{z_c^2}\right)^{3/2}.
\end{align}
For any star, $\rho=0$ at the surface (i.e. at $r=R$ or $\xi=\xi_1$). Therefore
at surface we have,
\begin{align}
 x_1=0, \,\,\,\,\, z_1=1, \,\,\,\,\, Q_1=\frac{1}{z_c} \,\,\,\,\, \text{at}
\,\,\,\,\, \xi=\xi_1. 
\end{align}
The expression of radius $R$ is,
\begin{align}
 R=a\xi_1=\sqrt{\frac{2 C_2}{\pi G}}\left(
\frac{1}{C_1 \mu z_c}\right) \xi_1
\end{align}
The total mass $M$ of the system is,
\begin{align}
 M =&\int_0^R 4 \pi r^2 \rho dr=4 \pi \left(C_1 \mu a^3 z_c^3\right)
\int_0^{Q_1} \xi^2 \left(Q^2-\frac{1}{z_c^2} \right)^{3/2} d\xi=4 \pi \left(C_1
\mu a^3 z_c^3\right) \left( -\xi^2 \frac{d Q}{ d\xi}\right)_1 \nonumber \\
=& \frac{4 \pi}{(C_1 \mu)^2}\left( \frac{2 C_2}{\pi G}\right)^{3/2} \left(
-\xi^2 \frac{d Q}{ d\xi}\right)_1
\end{align}
As before (\ref{DifferentialQ}) can be numerically solved for various values of
$\frac{1}{z_c}=0$ to $\frac{1}{z_c}=1$ (i.e. from $x_c=\infty$ to $x_c=0$). 
Let us consider the two extreme cases:
\begin{enumerate}
 \item {\bf $\bf \frac{1}{z_c}$=0}: This case corresponds to the fully
relativistic degenerate one. The numerical integration gives the values as
$\xi_1=6.89$ and $\left(-\xi^2 \frac{dQ}{d\xi}\right)_1=2.02$. Therefore, the
the corresponding values are $x_c=\infty$, $\rho_0=\infty$, $R=0$.
 \item {\bf $\bf \frac{1}{z_c}$=1}: This case corresponds to the 
nonrelativistic degenerate one. The numerical integration gives the values as
$\xi_1=\infty$ and $\left(-\xi^2 \frac{dQ}{d\xi}\right)_1=0$. In this case the
the values are $x_c=0$, $\rho_0=0$, $R=\infty$.
\end{enumerate}
The central density $\rho_0=C_1 \mu x_c$ 
decreases with decreasing $x_c$ i.e. the relativistic compact objects are denser
compared to the non-relativistic ones. Another point to note is that the radius
of the system decreases with increasing $x_c$, i.e. the massive white dwarfs are
smaller in size. 
\section{Luminosity of a white dwarf}
In this section we will explore the possible correction to the luminosity of a
white dwarf in DSR. Till now we saw the model of white dwarf where the whole white dwarf is assumed to be made up of degenerate gas, such that whole star is supported against the gravity crunch by the degenerate pressure. But we know that a star has a non-uniform density distribution i.e, the density goes to zero as radial distance becomes $R$ i.e, at the surface of the star. Hence the complete degenerate
description of white dwarf is inapplicable (see section 5.3 of reference \cite{paddy}, also see \cite{paddy1}\cite{stellar_astro}\cite{whitedwarf}). Therefore, we must consider the situation where the white dwarf is composed of partial degenerate matter. We wish to calculate the luminosity of a white dwarf using such a composition. We will start by describing the model for such a white dwarf. The detailed calculation for the luminosity will be shown thereafter. The assumptions involved will be stated explicitly.
\subsection{The model}
\begin{figure}
\centering
\includegraphics[width=0.7\textwidth]{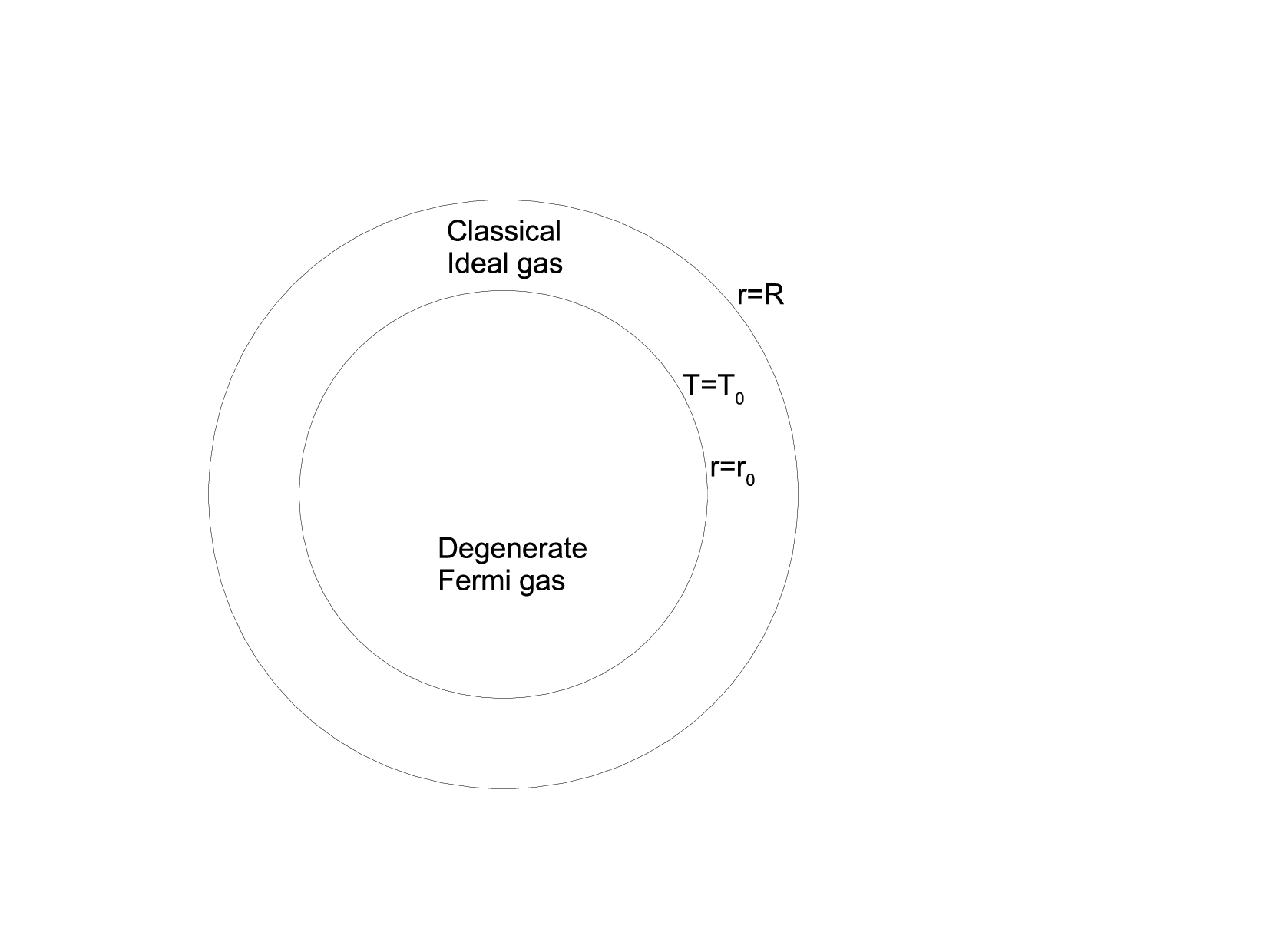}
\caption{Figure showing the partial degenerate model of white dwarf with
radiative envelope. We consider the
transition from quantum to classical is occurring at $r=r_0$ and $T=T_0$. Below
$r=r_0$ the gas is quantum and the pressure is degenerate pressure and above
$r=r_0$ we have the radiative envelope and the gas is classical ideal gas.
Therefore, above $r=r_0$ the
pressure is classical thermodynamic pressure $P=\frac{N_A}{\mu}\rho T$.}
\label{fig:luminos}
\end{figure}

As stated above our model is that of a partial degenerate matter constituting the white dwarf. What we mean is that upto certain radius $r_0$ from the centre the constituents of the white dwarf behave as degenerate gas and beyond that it behaves as non-degenerate matter. We have seen in the preceding sections that the degenerate pressure which holds the white dwarf against the gravity is calculated using the Fermi gas with $T=0$ K. On the other hand we know that the white dwarf star has some non zero finite temperature. 
To consider the radiative processes which leads us to understand the luminosity of the star we must have the non-degenerate radiative envelope. It is a standard observation that almost all the stars have a radiative envelope (see \cite{paddy}\cite{paddy1}\cite{stellar_astro}\cite{whitedwarf}\cite{stellar}). In our model we will also consider a non-degenerate radiative envelope encompassing the degenerate core of star such that up-to some $r_0$ from the centre of star we have degenerate Fermi gas and above which we have classical non-degenerate ideal gas till the surface. The model is shown in figure~\ref{fig:luminos}. The transition point $T=T_0$ represents the transition from quantum ideal degenerate Fermi gas to a classical ideal gas (note that this transition is smooth). Due to very high conductivity of the degenerate gas, the interior of the white dwarf upto the $r=r_0$ is isothermal with temperature $T_0$. The temperature gradient starts as we enter the non-degenerate part and it is this part that is responsible for the cooling of the white dwarf. As we move towards the surface of the star the temperature decreases making a finite surface temperature lower than $T_0$.
We will consider both the relativistic and non-relativistic cases for this
partial degenerate model in the DSR theory.
\subsection{The assumptions}
Usually modelling a star is very complicated and for many aspects of the star we still do not have a proper theory. But as is usually done we will take the standard set of assumptions and proceed further.
\begin{enumerate}
\item We have a spherically symmetric star in steady state with all the
physical variables depending only on radial coordinate $r$.
\item The radiative envelope is in local
thermal equilibrium, such that energy density is given by Planck's spectrum
$B_{\nu}$ corresponding to local temperature (see section 2.4 and exercise 2.9 of \cite{paddy} and also see section 9.2 of \cite{stellar_astro}).
\item The stellar fluid is stable against convection and so the entire flux is transferred through
radiative process only.
 \item Only one of the various processes of scattering are dominant making the polytrope to follow the power law for the opacity in terms of density and temperature (see section 2.4.1 of \cite{paddy}). The opacity $\gamma$ is given by $\gamma=\gamma_0 \rho^n T^{-s}$ (see section 2.2 of \cite{paddy}). The proportionality constant $\gamma_0$ depends on the composition of the star while the power indices $n$ and $s$ depend on the nature of the dominant scattering processes.
 \item Pressure $P$ follows ideal gas equation.
\end{enumerate}
\subsection{Variation of temperature and pressure and the expression of pressure in radiative envelope}
With the above model and assumptions in mind we will now attempt to calculate the luminosity of a white dwarf. As  was shown in \cite{Chandra:2016wth}, the energy density and pressure 
get a modification in DSR and so does the energy flux etc. 
We note that at such a local thermal equilibrium we have, for a given frequency,
the following relations
(see section 6.8 of \cite{paddy1})
\begin{enumerate}
 \item Intensity, $I_{\nu}\approx B_{\nu}=\frac{4\pi {\nu}^3}{e^{\frac{2\pi
\nu}{T}}-1}$
 \item Energy density, $U_{\nu}\approx 4\pi B_{\nu}$
 \item Luminosity, $L_{\nu}=\left(4\pi R^2\right)\frac{U_{\nu}}{4\pi} $
 \item Pressure, $P^{\alpha \beta}=P_{\nu} {\delta}^{\alpha}_{\beta}$
\end{enumerate}
Here $P_{\nu}$ is the radiation pressure for frequency $\nu$.
Under such an approximation the energy flux is given by (refer to equations (6.165)-(6.174) of \cite{paddy1}),
\begin{align}
 F^{\alpha}\approx \frac{-1}{\rho \gamma_{\nu}} \frac{\partial P^{\alpha
\beta}}{\partial x^{\beta}} \approx \frac{-1}{\rho \gamma_{\nu}}
\frac{\partial}{\partial x^{\beta}}(P_{\nu} {\delta}^{\alpha}_{\beta}),
\end{align}
where $\gamma_{\nu}$ is the opacity for a
particular frequency $\nu$ .
Note that the above equation is true for a frequency $\nu$. The actual equation is found by integrating over the frequency (refer to the \ref{energy_flux}).
Therefore, using the modified pressure and energy density
relation in \cite{Chandra:2016wth} and the derivatives of polylogarithm given by (see (4.1) in
\cite{polylogarithm}) $\frac{\partial}{\partial
\mu}[Li_n(e^{\mu})]=Li_{n-1}(e^{\mu})$, the required energy flux is given by (see section 2.2 of \cite{paddy}),
\begin{align}
F(r) &=\frac{L(r)}{4\pi r^2} =\left[\left(\frac{1}{ \rho \gamma_{\kappa}}\right)\left( \frac{-dP_{rad}}{dr}\right)\right]_{\text{boundary term}}+ \left[\left(\frac{1}{ \rho \gamma}\right) \left( \frac{-dP_{rad}}{dr}\right)\right] \nonumber \\
\end{align}
The expression for the boundary term is given by,
\begin{align}
\left(\frac{1}{3 \rho \gamma_{\kappa}}\right)
\left[\frac{{\kappa}^3}{{\pi}^2}
\ln\left(1-e^{-\frac{\kappa}{T}}\right)-\frac{{\kappa}^4}{{\pi}^2 T}
\frac{1}{e^{\frac{\kappa}{T}}-1}\right]\frac{dT}{dr}
\end{align}
and the other term is given by,
\begin{align} 
\left(\frac{1}{3\rho \gamma}\right)\left[-\frac{4{\pi}^2 T^3}{15}+\frac{24
T^3}{\pi^2} Li_4\left(e^{\frac{-\kappa}{T}}\right)+\frac{24 \kappa T^2 }{\pi^2}
Li_3\left(e^{\frac{-\kappa}{T}}\right)+\frac{12 {\kappa}^2
T}{\pi^2}Li_2\left(e^{\frac{-\kappa}{T}}\right)-\frac{4
{\kappa}^3}{\pi^2}\ln\left(1-e^{\frac{-\kappa}{T}}\right)+\frac{{\kappa}^4}{
\pi^2 T}
\frac{1}{e^{\frac{\kappa}{T}}-1}\right]\frac{dT}{dr}
\end{align}
Here $\gamma$ is the mean radiative opacity of the star, $\frac{1}{\gamma_{\kappa}}=\int_0^\kappa \frac{d \nu}{\gamma_{\nu}}$ is the mean radiative opacity corresponding to the boundary term (note that such a term is not present in SR case) and $\rho$ is the matter density of the star. The motivation for the
above expression is that if energy flux is $F(r)$ then momentum flux is also
$F(r)$ (as $c=1$), therefore the momentum scattered per second per unit volume
will be
$(n\sigma)F(r)=(\rho \gamma)F(r)$ (here $\sigma$ is the cross section and $n$ is
the number density of scatterers) and this is force per unit volume due to radiation on matter and is
related to $-\bigtriangledown P_{rad}$. 
The above equation can also be understood in terms of molecules in a room which are moving randomly but as soon as we open the windows we get a flow of air outside or inside depending on the pressure difference. Similarly, in a star the photons are moving randomly because of collisions. But since in a star the temperature decreases outward and the radiation pressure is smaller at greater distances from the center. This gradient in the radiation pressure is responsible for the net movement of photons toward the surface of the star that carries the radiative flux.

Assuming a fitting function for
$\gamma_{\kappa}$ and $\gamma(\rho ,T)$ exists, we can then invert to get,
\begin{align}\label{temperature_gradient}
 \frac{dT}{dr} &=\frac{L(r)}{4\pi r^2}\frac{3 \rho
\gamma_{\kappa}}{\left[\frac{{\kappa}^3}{{\pi}^2}
\ln\left(1-e^{-\frac{\kappa}{T}}\right)-\frac{{\kappa}^4}{{\pi}^2 T}
\frac{1}{e^{\frac{\kappa}{T}}-1}\right]} \nonumber \\
  &+\frac{L(r)}{4\pi r^2}\frac{3\rho \gamma}{\left[-\frac{4{\pi}^2
T^3}{15}+\frac{24 T^3}{\pi^2} Li_4\left(e^{\frac{-\kappa}{T}}\right)+\frac{24
\kappa T^2
}{\pi^2} Li_3\left(e^{\frac{-\kappa}{T}}\right)+\frac{12 {\kappa}^2
T}{\pi^2}Li_2\left(e^{\frac{-\kappa}{T}}\right)-\frac{4
{\kappa}^3}{\pi^2}\ln\left(1-e^{\frac{-\kappa}{T}}\right)+\frac{{\kappa}^4}{
\pi^2 T}
\frac{1}{e^{\frac{\kappa}{T}}-1}\right]}
\end{align}
Note that this should be applied locally for each $r$. Following assumption 3, we have assumed that
the fluid is stable against convection and so the entire flux is transferred through
radiative process only. Also according to assumption 4, there are several sources of opacity. The actual value depends on the medium and the various processes that are occurring at relevant
densities and temperatures.
As stated before, we will assume that the radiative envelope is in local
thermal equilibrium, such that energy density is given by Planck's spectrum
$B_{\nu}$ corresponding to local temperature.
Ignoring the convection we will first calculate the gradient as,
\begin{align}\label{gradient}
\bigtriangledown=\frac{d \ln T}{d \ln P} &=\frac{\frac{1}{T}\frac{dT}{dr}}{\frac{1}{P}\frac{dP}{dr}}=\frac{L(r)}{4\pi G M(r)}\frac{ 3 P
\gamma_{\kappa}}{\left[-\frac{T {\kappa}^3}{{\pi}^2}
\ln\left(1-e^{-\frac{\kappa}{T}}\right)+\frac{{\kappa}^4}{{\pi}^2}
\frac{1}{e^{\frac{\kappa}{T}}-1}\right]} \nonumber \\
  &+\frac{L(r)}{4\pi G M(r)}\frac{3 P \gamma}{\left[\frac{4{\pi}^2
T^4}{15}-\frac{24 T^4}{\pi^2} Li_4\left(e^{\frac{-\kappa}{T}}\right)-\frac{24
\kappa T^3
}{\pi^2} Li_3\left(e^{\frac{-\kappa}{T}}\right)-\frac{12 {\kappa}^2
T^2}{\pi^2}Li_2\left(e^{\frac{-\kappa}{T}}\right)+\frac{4 T
{\kappa}^3}{\pi^2}\ln\left(1-e^{\frac{-\kappa}{T}}\right)-\frac{{\kappa}^4}{
\pi^2}
\frac{1}{e^{\frac{\kappa}{T}}-1}\right]}
\end{align}
Here $P$ is the pressure of the fluid in the white dwarf, which we wish to calculate. Note that we have used equation (\ref{temperature_gradient}) and the equation of
hydrostatic equilibrium of a star given by,
\begin{align}
 \frac{d P}{d r}=-G\frac{M(r)\rho(r)}{r^2}
\end{align}
This is true for the spherically symmetric star in steady state with all the
physical variables depending only on radial coordinate $r$. Here $P(r)$,
$\rho(r)$ and $M(r)$ are the pressure, density at radius $r$ and mass contained
within a sphere of radius $r$.
The gas equation for non-degenerate gas is,
\begin{align}\label{ideal}
 P=\frac{N}{V} k_B T=\left(\frac{N m_H}{V
\rho}\right)\left(\frac{k_B}{m_H}\right)\rho T= \frac{N_A k_B \rho T}{\mu}
\end{align}
Here $\mu=\frac{V \rho}{N m_H}$ is the mean molecular weight,
$\frac{k_B}{m_H}=N_A k_B$ is the gas constant, $m_H$ is atomic mass constant,
$N_A$ is the Avogadro's Number.
Now using the dependence of opacity $\gamma=\gamma_0 \rho^n T^{-s}$ and equation (\ref{ideal}) we get,
\begin{align}
 & \frac{d T}{d P}=\frac{L}{4\pi G M}\frac{ 3 \gamma_{1 \kappa}
P^n}{\left[-\frac{T^{n+s} {\kappa}^3}{{\pi}^2}
\ln\left(1-e^{-\frac{\kappa}{T}}\right)+\frac{{\kappa}^4 T^{n+s-1}}{{\pi}^2}
\frac{1}{e^{\frac{\kappa}{T}}-1}\right]} \nonumber \\
  &+\frac{L}{4\pi G M}\frac{3 \gamma_1 P^n}{\left[\frac{4{\pi}^2
T^{n+s+3}}{15}-\frac{24 T^{n+s+3}}{\pi^2}
Li_4\left(e^{\frac{-\kappa}{T}}\right)-\frac{24
\kappa T^{n+s+2} }{\pi^2} Li_3\left(e^{\frac{-\kappa}{T}}\right)-\frac{12
{\kappa}^2
T^{n+s+1}}{\pi^2}Li_2\left(e^{\frac{-\kappa}{T}}\right)-\frac{4 {\kappa}^3
T^{n+s}}{\pi^2}Li_1\left(e^{\frac{-\kappa}{T}}\right)-\frac{{\kappa}^4
T^{n+s-1}}{\pi^2}
\frac{1}{e^{\frac{\kappa}{T}}-1}\right]}
\end{align}
Here $\gamma_1=\gamma_0 (\frac{\mu}{N_A})^n$ and $\gamma_{1\kappa}=\gamma_{0\kappa} (\frac{\mu}{N_A})^n$ and we have used $Li_1\left(e^{\frac{-\kappa}{T}}\right)=-\ln\left(1-e^{\frac{-\kappa}{T}}\right)$. 
Therefore we have,
\begin{align}\label{Pintegral}
 P^n dP &=\frac{4\pi G M}{3 L \gamma_{1 \kappa}} \left[-\frac{T^{n+s}
{\kappa}^3}{{\pi}^2} \ln\left(1-e^{-\frac{\kappa}{T}}\right)+\frac{{\kappa}^4
T^{n+s-1}}{{\pi}^2} \frac{1}{e^{\frac{\kappa}{T}}-1}\right] d T \nonumber \\
  &+\frac{4\pi G M}{3 L \gamma_1} \left[\frac{4{\pi}^2 T^{n+s+3}}{15}-\frac{24
T^{n+s+3}}{\pi^2} Li_4\left(e^{\frac{-\kappa}{T}}\right)-\frac{24 \kappa
T^{n+s+2} }{\pi^2}
Li_3\left(e^{\frac{-\kappa}{T}}\right)-\frac{12 {\kappa}^2
T^{n+s+1}}{\pi^2}Li_2\left(e^{\frac{-\kappa}{T}}\right)\right] d T \nonumber \\ 
  &+\frac{4\pi G M}{3 L \gamma_1} \left[\frac{4 {\kappa}^3
T^{n+s}}{\pi^2}\ln\left(1-e^{\frac{-\kappa}{T}}\right)-\frac{{\kappa}^4
T^{n+s-1}}{\pi^2}
\frac{1}{e^{\frac{\kappa}{T}}-1}\right] d T
\end{align}
Now we have to integrate this from $P_b$, $T_b$ (photospheric boundary
conditions) near the photosphere\footnote{Photosphere is the deepest region of a
luminous object, usually a star, that is transparent to photons of certain
wavelengths, in other words it
is the effective visual surface of the star. It is the region where the observed
optical photons originate.} to $P(r)$, $T(r)$ in the stellar envelope to get the
radiative polytrope equation such that $P(r)\geq P_b, T(r)\geq T_b$. We will now
use the expression of polylogarithm,
\begin{align}\label{poly}
Li_n(z)=\displaystyle{\sum_{a=1}^{\infty}} \frac{z^a}{a^n}, 
\end{align}
which is valid for $|z|<1$ (see (8.1) in \cite{polylogarithm}). This for our
case looks like $Li_n(e^{-\frac{\kappa}{T}})=\displaystyle{\sum_{a=1}^{\infty}}
\frac{e^{-\frac{a\kappa}{T}}}{a^n}$. We will try finding a general closed form
expression by considering,
\begin{align}\label{IntegralI}
 I= \int_{T_b}^{T(r)} \displaystyle{\sum_{j=1}^{\infty}} T^p
\frac{e^{-\frac{j\kappa}{T}}}{j^a} dT= \displaystyle{\sum_{j=1}^{\infty}}
\int_{T_b}^{T(r)} T^p \frac{e^{-\frac{j\kappa}{T}}}{j^a} dT,
\end{align}
provided $\displaystyle{\sum_{j=1}^{\infty}} \int_{T_b}^{T(r)} T^p
\frac{e^{-\frac{j\kappa}{T}}}{j^a} dT < \infty$. Now, making change of variables
as $x=\frac{j \kappa}{T}$ such that
$dT= -\frac{dx}{x^2}(j\kappa)$ and $\frac{j \kappa}{T_b}=\frac{j \kappa}{T_b},
\frac{j \kappa}{T(r)}=\frac{j \kappa}{T(r)}$, therefore we have
\begin{align}\label{IntegralGammaI}
 I=-\displaystyle{\sum_{j=1}^{\infty}} (j)^{p+1-a} (\kappa)^{p+1} \int_{\frac{j
\kappa}{T_b}}^{\frac{j \kappa}{T(r)}} \frac{e^{-x}}{x^{p+2}} dx &=
-(\kappa)^{p+1} \displaystyle{\sum_{j=1}^{\infty}} (j)^{p+1-a}
\left[\Gamma\left(-p-1, \frac{j \kappa}{T_b}\right)-\Gamma\left(-p-1, \frac{j
\kappa}{T(r)}\right) \right] \\
 &= -(\kappa)^{p+1} \displaystyle{\sum_{j=1}^{\infty}} (j)^{p+1-a}
\bigg[\Gamma\left(-p-1, \frac{j \kappa}{T_b}, \frac{j \kappa}{T(r)}\right)
\bigg]
\end{align}
Here $\Gamma(n, x_1,x_2)$ is the generalized incomplete gamma function defined
as $\Gamma(n, x_1,x_2)= \int_{x_1}^{x_2} \frac{e^{-t}dt}{t^{1-n}}=\Gamma(n,
x_1)-\Gamma(n, x_2)$ and $\Gamma(n, x)=\int_{x}^{\infty}
\frac{e^{-t}dt}{t^{1-n}}$ 
is the incomplete gamma function whose tabulated values are readily available or
can be numerically calculated for a given $\kappa$ and $T$. 
Remember this expression is true provided both $\frac{\kappa}{T_b} > 0$ and
$\frac{\kappa}{T(r)} >0$ (otherwise the integral diverges but this is true for
all physical cases) and, 
\begin{align}
(\kappa)^{p+1} \displaystyle{\sum_{j=1}^{\infty}} (j)^{p+1-a}
\bigg[\Gamma\left(-p-1, \frac{j \kappa}{T_b}\right)-\Gamma\left(-p-1, \frac{j
\kappa}{T(r)}\right) \bigg] < \infty
\end{align}
We can easily check by ratio test that this series is convergent given a
particular value of $\kappa$ and $T$ (Remember the value of incomplete gamma
function $\Gamma(n, x)$ decreases as the value of $x$ increases and therefore
converges very fast with increasing $x$ for a given $n$). Also in SR limit i.e.,
as $\kappa \rightarrow \infty$, for finite $T_b$ and $T(r)$, this whole term goes
to zero as expected. 
We can now proceed and write the closed form expression for the equation
(\ref{Pintegral}) as,
\begin{equation}
\begin{split}
P(r)&= \left[{P_b}^{n+1}+ (n+1)\left\{ \frac{4 G M}{3 \pi L \gamma_1}
\left[\frac{4{\pi}^4 {T_b}^{n+s+4}}{15(n+s+4)}-\frac{4{\pi}^4
{T(r)}^{n+s+4}}{15(n+s+4)} \right]
-\frac{4 G M(\kappa)^{n+s+4} }{3 \pi L \gamma_{1 \kappa}}
\left[\displaystyle{\sum_{j=1}^{\infty}} (j)^{n+s} \left[\Gamma\left(-n-s-1,
\frac{j \kappa}{T_b}, \frac{j \kappa}{T(r)}\right) \right]\right] \right.
\right. \\
&-\frac{4 G M(\kappa)^{n+s+4} }{3 \pi L \gamma_{1
\kappa}}\left[\displaystyle{\sum_{j=1}^{\infty}} (j)^{n+s}
\left[\Gamma\left(-n-s, \frac{j \kappa}{T_b}, \frac{j \kappa}{T(r)}\right)
\right]\right] 
+\frac{4 G M(\kappa)^{n+s+4} }{3 \pi L \gamma_1} \left[24
\displaystyle{\sum_{j=1}^{\infty}} (j)^{n+s} \left[\Gamma\left(-n-s-4, \frac{j
\kappa}{T_b}, \frac{j \kappa}{T(r)}\right) \right]\right]  \\ 
&+\frac{4 G M(\kappa)^{n+s+4} }{3 \pi L \gamma_1} \left[24
\displaystyle{\sum_{j=1}^{\infty}} (j)^{n+s} \left[\Gamma\left(-n-s-3, \frac{j
\kappa}{T_b}, \frac{j \kappa}{T(r)}\right) \right]\right] 
+\frac{4 G M(\kappa)^{n+s+4} }{3 \pi L \gamma_1} \left[12
\displaystyle{\sum_{j=1}^{\infty}} (j)^{n+s} \left[\Gamma\left(-n-s-2, \frac{j
\kappa}{T_b}, \frac{j \kappa}{T(r)}\right) \right]\right] \\ 
& \left. \left. +\frac{4 G M(\kappa)^{n+s+4}}{3 \pi L \gamma_1}\left[4
\displaystyle{\sum_{j=1}^{\infty}} (j)^{n+s} \left[\Gamma\left(-n-s-1, \frac{j
\kappa}{T_b}, \frac{j \kappa}{T(r)}\right) \right]\right] 
+\frac{4 G M(\kappa)^{n+s+4}}{3 \pi L \gamma_1}
\left[\displaystyle{\sum_{j=1}^{\infty}} (j)^{n+s} \left[\Gamma\left(-n-s,
\frac{j \kappa}{T_b}, \frac{j \kappa}{T(r)}\right) \right]\right]
\right\}\right]^{\frac{1}{n+1}}
\end{split}
\end{equation}
Using integration by parts of the incomplete gamma function we have a recurrence
relation as,
\begin{align}
 \Gamma(n+1,x)=n\Gamma(n,x)+x^n e^{-x}
\end{align}
Using above recurrence relation we can arrange the above expression in terms of one of the gammas. 
Further simplifying using the expression of the polylogarithm $Li_n(z)$ given in (\ref{poly}) we get the expression as (\ref{envelope}) given in appendix \ref{mod_P_L}.
The behaviour clearly depends mainly on the signs of $n+1$ and $n+s+4$. 
We can clearly see that the pressure of the DSR corrected partial degenerate gas is
smaller than the usual SR case for a given mass and temperature. 
This was also noted theoretically in
\cite{Camacho:2006qg}\cite{AmelinoCamelia:2009tv}.
We can
now express pressure in terms of the density and then equate it to density
obtained for degenerate Fermi gas expression, which we will see in the next
section.

\subsection{Calculation of the luminosity}
In this section we will try to find the actual expression for the luminosity of the white dwarf for both relativistic and the non-relativistic cases following the model described before shown in figure~\ref{fig:luminos}.
The transition point $T=T_0$ represents the transition from non-degenerate to degenerate matter and so $E_F=T_0$, we now introduce
$\mu_e=\frac{\rho}{nm_p}$ (where $m_p$ is mass of proton and $\mu_e$ is the mass
per electron) and substitute for $n$ in (\ref{p_F}) to get,
\begin{equation}
  p_F = (3\pi^2)^{1/3}\left(\frac{\rho}{\mu_e m_p}\right)^{1/3}
\end{equation}
which therefore gives,
\begin{equation}
   \frac{\rho}{\mu_e}=\frac{m_p}{3\pi^2}(p_F)^{3}
   \label{density}
\end{equation}
We will start by considering the relativistic case first.
\subsection*{Relativistic case:}
We will first consider the relativistic case with dispersion relation (\ref{MS})
giving ${p_F}^2={E_F}^2-m^2\left(1-\frac{E_F}{\kappa}\right)^2$. Substituting
the value of $p_F$ and $E_F=T_0$ in (\ref{density}),
\begin{equation}\label{density_reldenenerate}
 \rho_0=\frac{m_p\mu_e}{3\pi^2}
\left[{T_0}^2-m^2\left(1-\frac{T_0}{\kappa}\right)^2\right]^{\frac{3}{2}}
\end{equation}
Here $\rho_0$ is the mass density different from
energy density calculated in \cite{Chandra:2011nj}.
Note that as stated before, the region $r<r_0$ is constant temperature ($T_0$) region because of
very high conductivity of electrons. This is the region of highly dense and degenerate
electron gas with long mean free path and therefore the whole matter upto $r_0$ is isothermal with constant temperature $T_0$.
The non-degenerate envelope is mainly responsible for the luminosity of the white dwarf
(see discussion in section 5.3 of \cite{paddy}). 
We will express the envelope pressure expression (\ref{envelope}) in terms of density using (\ref{ideal}) and equate the corresponding density expression to   
(\ref{density_reldenenerate}), to get the expression
of luminosity $L$ of relativistic gas for particular $n$ and $s$ and the chosen
boundary conditions $T_b$ and $P_b$ as (\ref{luminosity_rel}) given in appendix \ref{mod_P_L}. Since luminosity is proportional to the energy flux and for the radiative envelope we put an ultraviolet cut-off on the maximum single particle energy. Therefore, the flux and so the luminosity gets a negative correction. Because of the same reason the luminosity expression is nonperturbative in the SR limit. This gives the expected standard result in the SR limit. 
\subsection*{Non-relativistic case:}
To get the result for the non-relativistic (which we will call DNR) case we
first
need to know the non-relativistic limit of DSR. Normally, to take the NR limit of
SR we write the SR dispersion relation
and consider $p<<m_0$ and expand the relation in this limit. We will do the
same procedure of considering the DSR dispersion relation and expanding it in
the limit $p<<m_0$.
Although there are references
as \cite{Coraddu:2009sb} and \cite{Jafari:2011dd} which discuss the DNR limit of
DSR, but they both take the expansion in the limit $p<<m$. It seems natural to
make the
expansion in the limit $p<<m_0$ as $m_0$ is the rest mass. 
On the other hand the invariant mass $m$ has no such physical meaning (not to be confused with the word `\emph{invariant}` as the rest mass $m_0$ is also invariant under a DSR transformation). 
Are these limits equivalent?
The answer to this question is no, as a case may arise when $\frac{p}{m}<<1$ but $\frac{p}{m_0}\approx1$, making the two limits different from each other. It is so when $p
\approx \kappa$, $m_0 \approx \kappa$ $\Leftrightarrow$ $m \rightarrow
\infty$.
We will, therefore, start with the dispersion relation in (\ref{MS}) and
rearrange
the terms and complete the squares to get, 
\begin{equation}
 E=\frac{\frac{-m^2}{\kappa^2}\pm\sqrt{m^2+p^2\left(1-\frac{m^2}{\kappa^2}
\right)}}{1-\frac
{m^2}{\kappa^2}}
\end{equation}
Since we want the expansion in limit $p<<m_0$, we substitute $m$ by $m_0$
using (\ref{m}) to get
\begin{equation}
  E=\left(\frac{-{m_0}^2}{2m_0-\kappa}\right) \pm
\frac{m_0(m_0-\kappa)}{2m_0-\kappa}\sqrt{1+\frac{p^2}{{m_0}^2}\left(\frac{
\kappa-2m_0
}{\kappa}\right)}
\end{equation}
We then make the following assumptions before doing the binomial expansion,
\subsection*{Assumptions:}
\begin{enumerate}
 \item $\frac{p}{m_0}<<1$. Note that this is the DNR assumption we expect
physically.
 \item $m_0$, $\kappa$ both are finite. Remember that in case of DSR this
assumption
is valid from the way it has originally been formulated ($0\leq m_0
\leq\kappa$).
\end{enumerate}
With above assumptions in mind we do the expansion and keep the first order
terms in $\frac{p}{m_0}$ to get,
\begin{equation}
  E \simeq \left(\frac{-{m_0}^2}{2m_0-\kappa}\right) \pm
\frac{m_0(m_0-\kappa)}{2m_0-\kappa}\left[1+\frac{p^2}{2{m_0}^2}\left(\frac{
\kappa-2m_0}{\kappa}\right)\right].
\end{equation}
Considering the positive value first we get from further simplification,
\begin{equation}
 E=m_0+\frac{p^2}{2m_0}\left[1-\frac{m_0}{\kappa}\right]=m_0 + \frac{p^2}{2m}
\end{equation}
Here $m_0$ is the rest mass as expected and $m$ is the non-relativistic inertial
mass. Our result matches with the literature \cite{Coraddu:2009sb} and
\cite{Jafari:2011dd}. Let us now consider the negative value,
\begin{equation}
 E=\frac{m_0\kappa}{2m_0-\kappa}-\frac{p^2}{2m}=-\left[\frac{m_0\kappa}{
\kappa-2m_0}+\frac{p^2}{2m}\right]
\end{equation}
These two energies correspond to the particle and the anti-particle
respectively.
Note that the rest mass of the particle and the anti-particle are different.
But, DNR dynamics
depends on $m$ which is the same for both.
Substituting the value of $p_F$ and $E_F=T_0$ in (\ref{density}) we get,
\begin{equation}\label{density_nonreldenenerate}
 \rho_0=\frac{m_p\mu_e}{3\pi^2} \left(2mT_0\right)^{\frac{3}{2}}
\end{equation}
As done before we will express the envelope pressure expression (\ref{envelope}) in terms of density using (\ref{ideal}) and equate the corresponding density expression to the density above at $T=T_0$ to get the relation between $M$, $L$ for non-relativistic gas for given $n$, $s$ and chosen boundary
conditions on $T_b$ and $P_b$ as (\ref{luminosity_nonrel}) given in appendix \ref{mod_P_L}. 
In this case also the luminosity gets a negative correction. The luminosity expression is as expected nonperturbative in the SR limit. This expression also gives the expected standard result in the SR limit. 
So the luminosity in both the cases are lower than the usual value and nonperturbative in the SR limit.
\section{Summary and Future Works}
In this paper we started with the study of the effect of a relativistically invariant energy scale on the thermodynamics of a degenerate Fermi gas. 
We considered the model of Modified Dispersion Relation(MDR) with an invariant ultraviolet cut-off on the single particle energy. 
We found the correction to the thermodynamic pressure and the total energy of the degenerate Fermi gas in all the three cases $m<\kappa$, $m=\kappa$ and $m>\kappa$.
We discussed the number density $n$ and mass $m$ dependence of the degenerate pressure for $m<\kappa$ case. 
We found that the degenerate pressure is perturbative in the SR limit, a result unexpected for the theory with an ultraviolet cut-off. 
We also, briefly discussed the two extreme
nonrelativistic and the ultrarelativistic limits of the pressure. We took the white dwarf stars as an example and studied its modified dynamics.
For the usual particle i.e, in case of $m<\kappa$, as is obvious from figure~\ref{fig:Pplot}, we found that the equilibrium degeneracy pressure is greater than, equal to 
and less than the SR value for particular masses of the considered compact object such as white dwarf. Also as shown in figure~\ref{fig:MRplot}, for given masses the value of the radius of the white dwarf is found to be less than, equal to and greater than the usual SR value. Since energy density is related directly to mass density so
denser compact objects are expected to show better measurable correction due to such a modification.
We do not get the Chandrasekhar limit in the other two cases as expected.
The Chandrasekhar mass limit for a white dwarf in this case is greater than the usual SR value which is clearly visible from the plot. One of the major predictions of our theory is that it makes an attempt to explain the observed lower radius white dwarfs and also predicts the white dwarfs having radius greater than that predicted by the present SR theory.
The correction (see (\ref{chandra_correct})) is purely perturbative in the SR limit which is quite unusual for a theory having an ultraviolet energy cut-off. Therefore we conclude that this correction is solely because of the modified dispersion relation. The other two cases $m=\kappa$ and $m>\kappa$ has also been studied where we do not get any limit on the white dwarf mass. Note that the presence of observed white dwarfs having radius lesser than the SR case may find an explanation if they are modelled using a modified dispersion relation.
This result has also been found using the modified Lane-Emden equation assuming radial density distribution. General modified structure equation has also been discussed in detail.
Along with this it was shown in \cite{Chandra:2016wth} that the Stefan-Boltzmann
law gets modified in DSR and so does the luminosity. We therefore calculate the
the luminosity of such a white dwarf in DSR both in
nonrelativistic and relativistic cases.
We noted that since the correction in pressure is negative for a given mass and temperature and so is the
correction in the luminosity as well. 
The correction to the luminosity of a white dwarf is nonperturbative in SR limit as expected because of the presence of an ultraviolet energy cut-off.
In future, one can also do similar analysis of other dense and compact stars like neutron stars etc. 
Since neutron stars are denser than the white dwarfs, one expects more prominent signature to the DSR correction.
Given the modification of degenerate fermions in the formalism discussed in this article, one can further explore the general modifications in the thermodynamics of a fermion gas (not necessarily degenerate). To study the thermodynamics of a black hole one needs to formulate the DSR on a curved spacetime.
\section{Acknowledgement}

DKM would like to thank Vinay Vaibhav, Prashanth Raman, Anirban Karan and Issan Patri for many
useful suggestions. NC and DKM would like to thank Sandeep Chatterjee for many
discussions.

\begin{appendices}
\section{Energy flux for radiative processes}\label{energy_flux}
We have the expression of the energy flux given by
\begin{align}
 F^{\alpha}\approx \frac{-1}{\rho \gamma_{\nu}} \frac{\partial P^{\alpha
\beta}}{\partial x^{\beta}} \approx \frac{-1}{\rho \gamma_{\nu}}
\frac{\partial}{\partial x^{\beta}}(P_{\nu} {\delta}^{\alpha}_{\beta})
\end{align}
Integrating over frequency and using the modified pressure and energy density
relation in \cite{Chandra:2016wth} we get,
\begin{align}
 {\bf{F}}_{rad} &= \left(\frac{1}{3 \rho \gamma_{\kappa}}\right)
\bigtriangledown
\left[\frac{T{\kappa}^3}{{\pi}^2} \ln\left(1-e^{-\frac{\kappa}{T}}\right)\right]
\nonumber
\\
 &- \left(\frac{1}{3\rho \gamma_{R}}\right) \bigtriangledown \left[\frac{{\pi}^2
T^4}{15}-\left[\left(\frac{6T^4}{\pi^2}\right)Li_4\left(e^{-\frac{\kappa}{T}}
\right)+\left(\frac{6\kappa T^3}{\pi^2}\right)
Li_3\left(e^{-\frac{\kappa}{T}}\right)+\left(\frac{3{\kappa}^2
T^2}{\pi^2}\right) 
Li_2\left(e^{-\frac{\kappa}{T}}\right)-\left(\frac{{\kappa}^3 T}{\pi^2}\right)
\ln\left(1-e^{-\frac{\kappa}{T}}\right)\right]\right]
\end{align}
This relation can be used to find the radiative force on per unit volume of matter using equation (6.167) of \cite{paddy1}.
Here $\bf{F}_{rad}$ is the total radiative flux and
$\frac{1}{\gamma_{\kappa}}=\int_0^\kappa \frac{d \nu}{\gamma_{\nu}}$, 
$\gamma_{R}$ is the
modified Rosseland mean opacity defined as,
\begin{align}
 \frac{1}{\gamma_{R}}=\frac{\int_0^{\kappa} \frac{1}{\gamma_{\nu}}
\left(\frac{\partial B_{\nu}}{\partial T}\right) d\nu}{\int_0^{\kappa}
\left(\frac{\partial
B_{\nu}}{\partial T}\right) d\nu}
\end{align}
where we have used the following relation,
\begin{align}
 \int_0^{\kappa} \left(\frac{\partial B_{\nu}}{\partial T}\right) d\nu &=
\frac{\partial}{{\partial T}} \int_0^{\kappa} (B_{\nu}) d\nu \nonumber \\
  &= \frac{1}{4\pi} \frac{\partial}{{\partial T}}\left[\frac{{\pi}^2
T^4}{15}-\left[\left(\frac{6T^4}{\pi^2}\right)Li_4\left(e^{-\frac{\kappa}{T}}
\right)+\left(\frac{6\kappa T^3}{\pi^2}\right)
Li_3\left(e^{-\frac{\kappa}{T}}\right)+\left(\frac{3{\kappa}^2
T^2}{\pi^2}\right) 
Li_2\left(e^{-\frac{\kappa}{T}}\right)-\left(\frac{{\kappa}^3 T}{\pi^2}\right)
\ln\left(1-e^{-\frac{\kappa}{T}}\right)\right]\right]
\end{align}
This expression in general can be written by considering the mean opacity as $\gamma$ instead of $\gamma_R$. Now assuming the physical quantities depend on distance $r$, the vector equation reduce to the scalar one given in the text above.
\section{Modified Pressure and Luminosity Expressions}\label{mod_P_L}
The expression of the pressure becomes,
\begin{equation}\label{envelope}
\begin{split}
P(r)&= \left[{P_b}^{n+1}+ (n+1)\left\{ \frac{4 G M}{3 \pi L \gamma_1}
\left[\frac{4{\pi}^4 {T_b}^{n+s+4}}{15(n+s+4)}-\frac{4{\pi}^4
{T(r)}^{n+s+4}}{15(n+s+4)} \right] \right. \right. \\
&-\frac{4 G M(\kappa)^{n+s+4} }{3 \pi L \gamma_{1 \kappa}} \left[-\left( Li_4
\left(e^{\frac{-\kappa}{T_b}}\right) \left(\frac{T_b}{\kappa} \right)^{n+s+4}-
Li_4 \left(e^{\frac{-\kappa}{T(r)}}\right)\left(\frac{T(r)}{\kappa}
\right)^{n+s+4}\right) (n+s)(n+s+2)(n+s+3) \right.  \\
&+\left( Li_3 \left(e^{\frac{-\kappa}{T_b}}\right)\left(\frac{T_b}{\kappa}
\right)^{n+s+3}- Li_3
\left(e^{\frac{-\kappa}{T(r)}}\right)\left(\frac{T(r)}{\kappa}
\right)^{n+s+3}\right) (n+s)(n+s+2) \\
&-\left(  Li_2 \left(e^{\frac{-\kappa}{T_b}}\right)\left(\frac{T_b}{\kappa}
\right)^{n+s+2}- Li_2 \left(e^{\frac{-\kappa}{T(r)}}\right)
\left(\frac{T(r)}{\kappa} \right)^{n+s+2}\right) (n+s) \\
&+\left(Li_1 \left(e^{\frac{-\kappa}{T_b}}\right)\left(\frac{T_(r)}{\kappa}
\right)^{n+s+1}-Li_1
\left(e^{\frac{-\kappa}{T(r)}}\right)\left(\frac{T(r)}{\kappa}
\right)^{n+s+1}\right)  \\
&\left. + \displaystyle{\sum_{j=1}^{\infty}} j^{n+s} \Gamma\left(-n-s-4,\frac{j
\kappa}{T_b},\frac{j \kappa}{T(r)}\right) (n+s)(n+s+2)(n+s+3)(n+s+4)\right]
\\
&+\frac{4 G M(\kappa)^{n+s+4} }{3 \pi L \gamma_1} \left[-\left( Li_4
\left(e^{\frac{-\kappa}{T_b}}\right)\left(\frac{T_b}{\kappa}
\right)^{n+s+4}-Li_4
\left(e^{\frac{-\kappa}{T(r)}}\right)\left(\frac{T(r)}{\kappa}
\right)^{n+s+4}\right) (n+s-1)[6+n^2+s(s+3)+n(2s+3)] \right. \\
&+\left(Li_3 \left(e^{\frac{-\kappa}{T_b}}\right)\left(\frac{T_b}{\kappa}
\right)^{n+s+3}-Li_3
\left(e^{\frac{-\kappa}{T(r)}}\right)\left(\frac{T(r)}{\kappa}
\right)^{n+s+3}\right)[6+n^2+s(s-1)+n(2s-1)]  \\
&-\left(Li_2 \left(e^{\frac{-\kappa}{T_b}}\right)\left(\frac{T_b}{\kappa}
\right)^{n+s+2}-Li_2
\left(e^{\frac{-\kappa}{T(r)}}\right)\left(\frac{T(r)}{\kappa}
\right)^{n+s+2}\right) (n+s-3)  \\
&+\left(Li_1 \left(e^{\frac{-\kappa}{T_b}}\right)\left(\frac{T_b}{\kappa}
\right)^{n+s+1}-Li_1
\left(e^{\frac{-\kappa}{T(r)}}\right)\left(\frac{T(r)}{\kappa}
\right)^{n+s+1}\right)  \\
& \left. \left. \left. + \displaystyle{\sum_{j=1}^{\infty}} j^{n+s}
\Gamma\left(-n-s-4,\frac{j
\kappa}{T_b},\frac{j \kappa}{T(r)}\right) (n+s)(n+s+1)(n+s+2)(n+s+3)\right]
\right\}\right]^{\frac{1}{n+1}}.
\end{split}
\end{equation}

The corresponding expression of luminosity in relativistic case is,
\begin{equation}\label{luminosity_rel}
\begin{split}
L&=\frac{(n+1)}{P^{\prime}}\left\{ \frac{4 G M}{3 \pi L \gamma_1}
\left[\frac{4{\pi}^4 {T_b}^{n+s+4}}{15(n+s+4)}-\frac{4{\pi}^4
{T(r)}^{n+s+4}}{15(n+s+4)} \right]  \right. \\
&-\frac{4 G M(\kappa)^{n+s+4} }{3 \pi L \gamma_{1 \kappa}} \left[-\left( Li_4
\left(e^{\frac{-\kappa}{T_b}}\right) \left(\frac{T_b}{\kappa} \right)^{n+s+4}-
Li_4 \left(e^{\frac{-\kappa}{T(r)}}\right)\left(\frac{T(r)}{\kappa}
\right)^{n+s+4}\right) (n+s)(n+s+2)(n+s+3) \right.  \\
&+\left( Li_3 \left(e^{\frac{-\kappa}{T_b}}\right)\left(\frac{T_b}{\kappa}
\right)^{n+s+3}- Li_3
\left(e^{\frac{-\kappa}{T(r)}}\right)\left(\frac{T(r)}{\kappa}
\right)^{n+s+3}\right) (n+s)(n+s+2) \\
&-\left(  Li_2 \left(e^{\frac{-\kappa}{T_b}}\right)\left(\frac{T_b}{\kappa}
\right)^{n+s+2}- Li_2 \left(e^{\frac{-\kappa}{T(r)}}\right)
\left(\frac{T(r)}{\kappa} \right)^{n+s+2}\right) (n+s)
+\left(Li_1 \left(e^{\frac{-\kappa}{T_b}}\right)\left(\frac{T_(r)}{\kappa}
\right)^{n+s+1}-Li_1
\left(e^{\frac{-\kappa}{T(r)}}\right)\left(\frac{T(r)}{\kappa}
\right)^{n+s+1}\right)  \\
&\left. + \displaystyle{\sum_{j=1}^{\infty}} j^{n+s} \Gamma\left(-n-s-4,\frac{j
\kappa}{T_b},\frac{j \kappa}{T(r)}\right) (n+s)(n+s+2)(n+s+3)(n+s+4)\right]
\\
&+\frac{4 G M(\kappa)^{n+s+4} }{3 \pi L \gamma_1} \left[-\left( Li_4
\left(e^{\frac{-\kappa}{T_b}}\right)\left(\frac{T_b}{\kappa}
\right)^{n+s+4}-Li_4
\left(e^{\frac{-\kappa}{T(r)}}\right)\left(\frac{T(r)}{\kappa}
\right)^{n+s+4}\right) (n+s-1)[6+n^2+s(s+3)+n(2s+3)] \right. \\
&+\left(Li_3 \left(e^{\frac{-\kappa}{T_b}}\right)\left(\frac{T_b}{\kappa}
\right)^{n+s+3}-Li_3
\left(e^{\frac{-\kappa}{T(r)}}\right)\left(\frac{T(r)}{\kappa}
\right)^{n+s+3}\right)[6+n^2+s(s-1)+n(2s-1)]  \\
&-\left(Li_2 \left(e^{\frac{-\kappa}{T_b}}\right)\left(\frac{T_b}{\kappa}
\right)^{n+s+2}-Li_2
\left(e^{\frac{-\kappa}{T(r)}}\right)\left(\frac{T(r)}{\kappa}
\right)^{n+s+2}\right) (n+s-3) 
+\left(Li_1 \left(e^{\frac{-\kappa}{T_b}}\right)\left(\frac{T_b}{\kappa}
\right)^{n+s+1}-Li_1
\left(e^{\frac{-\kappa}{T(r)}}\right)\left(\frac{T(r)}{\kappa}
\right)^{n+s+1}\right)  \\
& \left. \left. + \displaystyle{\sum_{j=1}^{\infty}} j^{n+s}
\Gamma\left(-n-s-4,\frac{j
\kappa}{T_b},\frac{j \kappa}{T(r)}\right) (n+s)(n+s+1)(n+s+2)(n+s+3)\right]
\right\},
\end{split}
\end{equation}
where $P^{\prime}=\left[\left(\frac{N_A T_0 m_p \mu_e
\left[{T_0}^2-m^2\left(1-\frac{T_0}{\kappa}\right)^2\right]^{\frac{3}{2}} }{3
\pi^2 \mu}\right)^{n+1}-P_b^{n+1}\right]$.

The expression of luminosity in nonrelativistic case is,
\begin{equation}\label{luminosity_nonrel}
\begin{split}
L&=\frac{(n+1)}{P^{\prime}}\left\{ \frac{4 G M}{3 \pi L \gamma_1}
\left[\frac{4{\pi}^4 {T_b}^{n+s+4}}{15(n+s+4)}-\frac{4{\pi}^4
{T(r)}^{n+s+4}}{15(n+s+4)} \right]  \right. \\
&-\frac{4 G M(\kappa)^{n+s+4} }{3 \pi L \gamma_{1 \kappa}} \left[-\left( Li_4
\left(e^{\frac{-\kappa}{T_b}}\right) \left(\frac{T_b}{\kappa} \right)^{n+s+4}-
Li_4 \left(e^{\frac{-\kappa}{T(r)}}\right)\left(\frac{T(r)}{\kappa}
\right)^{n+s+4}\right) (n+s)(n+s+2)(n+s+3) \right.  \\
&+\left( Li_3 \left(e^{\frac{-\kappa}{T_b}}\right)\left(\frac{T_b}{\kappa}
\right)^{n+s+3}- Li_3
\left(e^{\frac{-\kappa}{T(r)}}\right)\left(\frac{T(r)}{\kappa}
\right)^{n+s+3}\right) (n+s)(n+s+2) \\
&-\left(  Li_2 \left(e^{\frac{-\kappa}{T_b}}\right)\left(\frac{T_b}{\kappa}
\right)^{n+s+2}- Li_2 \left(e^{\frac{-\kappa}{T(r)}}\right)
\left(\frac{T(r)}{\kappa} \right)^{n+s+2}\right) (n+s)
+\left(Li_1 \left(e^{\frac{-\kappa}{T_b}}\right)\left(\frac{T_(r)}{\kappa}
\right)^{n+s+1}-Li_1
\left(e^{\frac{-\kappa}{T(r)}}\right)\left(\frac{T(r)}{\kappa}
\right)^{n+s+1}\right)  \\
&\left. + \displaystyle{\sum_{j=1}^{\infty}} j^{n+s} \Gamma\left(-n-s-4,\frac{j
\kappa}{T_b},\frac{j \kappa}{T(r)}\right) (n+s)(n+s+2)(n+s+3)(n+s+4)\right]
\\
&+\frac{4 G M(\kappa)^{n+s+4} }{3 \pi L \gamma_1} \left[-\left( Li_4
\left(e^{\frac{-\kappa}{T_b}}\right)\left(\frac{T_b}{\kappa}
\right)^{n+s+4}-Li_4
\left(e^{\frac{-\kappa}{T(r)}}\right)\left(\frac{T(r)}{\kappa}
\right)^{n+s+4}\right) (n+s-1)[6+n^2+s(s+3)+n(2s+3)] \right. \\
&+\left(Li_3 \left(e^{\frac{-\kappa}{T_b}}\right)\left(\frac{T_b}{\kappa}
\right)^{n+s+3}-Li_3
\left(e^{\frac{-\kappa}{T(r)}}\right)\left(\frac{T(r)}{\kappa}
\right)^{n+s+3}\right)[6+n^2+s(s-1)+n(2s-1)]  \\
&-\left(Li_2 \left(e^{\frac{-\kappa}{T_b}}\right)\left(\frac{T_b}{\kappa}
\right)^{n+s+2}-Li_2
\left(e^{\frac{-\kappa}{T(r)}}\right)\left(\frac{T(r)}{\kappa}
\right)^{n+s+2}\right) (n+s-3) 
+\left(Li_1 \left(e^{\frac{-\kappa}{T_b}}\right)\left(\frac{T_b}{\kappa}
\right)^{n+s+1}-Li_1
\left(e^{\frac{-\kappa}{T(r)}}\right)\left(\frac{T(r)}{\kappa}
\right)^{n+s+1}\right)  \\
& \left. \left. + \displaystyle{\sum_{j=1}^{\infty}} j^{n+s}
\Gamma\left(-n-s-4,\frac{j
\kappa}{T_b},\frac{j \kappa}{T(r)}\right) (n+s)(n+s+1)(n+s+2)(n+s+3)\right]
\right\},
\end{split}
\end{equation}
where $P^{\prime}=\left[\left(\frac{N_A T_0 m_p \mu_e
\left[2 m T_0 \right]^{\frac{3}{2}} }{3
\pi^2 \mu}\right)^{n+1}-P_b^{n+1}\right]$.

\end{appendices}

\end{document}